\documentclass{aa}

\usepackage{graphicx}
\usepackage{natbib}

\usepackage{txfonts}
\usepackage{color}
\usepackage{longtable}

\newcommand{\code}[1]{\texttt{#1}}
  
\usepackage{hyperref}
\begin{document}

   \title{Effects of Multiphase Gas and Projection on X-ray Observables in Simulated Galaxy Clusters as Seen by \textit{eROSITA}}

   \author{J. ZuHone \inst{1}, 
   Y. E. Bahar \inst{2}, 
   V. Biffi \inst{3,4}, 
   K. Dolag \inst{5,6}, 
   J. Sanders \inst{2}, 
   E. Bulbul \inst{2}, 
   T. Liu \inst{2},
   T. Dauser\inst{7}, 
   O. K{\"o}nig\inst{7},
   X. Zhang\inst{2}, \and
   V. Ghirardini\inst{2}
   } 
   \institute{Center for Astrophysics | Harvard and Smithsonian, 60 Garden St., Cambridge, MA 02138,
              USA\\
              \email{john.zuhone@cfa.harvard.edu}
              \and
              Max-Planck-Institut f{\"u}r Extraterrestrische Physik (MPE), Gie{\ss}enbachstra{\ss}e
              1, D-85748 Garching bei M{\"u}nchen, Germany
              \and
              INAF --- Osservatorio Astronomico di Trieste, via Tiepolo 11, 34143 Trieste, Italy
              \and
              IFPU --- Institute for Fundamental Physics of the Universe, Via Beirut 2, 34014
              Trieste, Italy 
              \and
              Universit\"ats-Sternwarte, Fakult\"at f\"ur Physik, Ludwig-Maximilians-Universit\"at M\"unchen, Scheinerstr.1, 81679 M\"unchen, Germany 
              \and
              Max-Planck-Institut f{\"u}r Astrophysik (MPA), Karl-Schwarzschild-Stra{\ss}e 1,
              D-85748 Garching bei M{\"u}nchen, Germany 
              \and
                Remeis-Observatory \& ECAP, FAU Erlangen-N\"urnberg, Sternwartstr. 7, 96049 Bamberg, Germany
              }

   \date{Received XXX; accepted XXX}

 
  \abstract
   {Being the most massive bound objects in the recent history of the universe, the number density of galaxy clusters as a function of mass and redshift is a sensitive function of the cosmological parameters. To use clusters for cosmological parameter studies, it is necessary to determine their masses as accurately as possible, which is typically done via scaling relations between mass and observables.}
   {X-ray observables can be biased by a number of effects, including multiphase gas and projection effects, especially in the case where cluster temperatures and luminosities are estimated from single-model fits to all of the emission with an overdensity radius such as $r_{500c}$. Using simulated galaxy clusters from a realistic cosmological simulation, we seek to determine the importance of these biases in the context of \textit{Spectrum-Roentgen-Gamma/eROSITA} observations of clusters.}
   {We extract clusters from the \code{Box2\_hr} simulation from the Magneticum suite, and simulate
    synthetic \textit{eROSITA} observations of these clusters using \code{PHOX} to generate the
    photons and the end-to-end simulator \code{SIXTE} to trace them through the optics and simulate the detection process.  We fit the spectra from
    these observations and compare the fitted temperatures and luminosities to the quantities
    derived from the simulations. We fitted an intrinsically scattered $L_{\rm X}-T$ scaling relation to these measurements following a Bayesian approach with which we fully took into account the selection effects and the mass function.}
   {The largest biases on the estimated temperature and luminosities of the clusters come from the inadequacy of single-temperature model fits to represent emission from multiphase gas, as well as a bias arising from cluster emission within the \textit{projected} $r_{500c}$ along the line of sight but outside of the \textit{spherical} $r_{500c}$. We find that the biases on temperature and luminosity due to the projection of emission from other clusters within $r_{500c}$ is comparatively small. We find eROSITA-like measurements of Magneticum clusters following a $L_{\rm X}-T$ scaling relation that has a broadly consistent but slightly shallower slope compared to the literature. We also find that the intrinsic scatter of $L_{\rm X}$ at given $T$ is lower compared to the recent observational results where the selection effects are fully considered.}
   {}

   \keywords{galaxy: clusters: intracluster medium –- method: numerical –- X-ray: galaxies: cluster}

   \titlerunning{Properties of Magneticum Clusters Observed by eROSITA}
   \authorrunning{J. ZuHone et al.}

   \maketitle
%

\section{Introduction}

Galaxy clusters are the natural endpoints of the process of hierarchical structure formation in a
$\Lambda$CDM universe at the current epoch. Given their size, galaxy clusters are representative of
the material properties of the universe as a whole. The mass budget of clusters is dominated by dark
matter (DM), at roughly $\sim$80--90\% of the total mass, with the remaining $\sim$10--20\%
comprised of baryons. Of these, the vast majority reside in the diffuse ($n \sim
10^{-4}-10^{-1}$~cm$^{-3}$) and hot ($kT \sim 1-10$~keV) ionized plasma known as the intracluster
medium (ICM), that emits in the X-ray band and is visible at mm wavelengths via the
Sunyaev-Zeldovich (SZ) effect. The stars in the galaxies and the ``intracluster light'' of stars
outside of galaxies only comprise a few percent by mass. 

Clusters of galaxies are important probes of cosmology, due to the fact that their number density as
a function of mass and redshift is sensitive to the values of the cosmological parameters. This
requires accurate mass measurements for clusters. Gravitational lensing can be used to estimate
masses directly in a number of systems, but most clusters do not exhibit sufficient gravitational
lensing to produce well-constrained mass models \citep{Ramos-Ceja2022, Chiu2022}. Thus, like most
observed structures in the universe, the masses of galaxy clusters must typically be inferred from
the kinematics of the luminous material, in this case the ICM under the assumption of hydrostatic
equilibrium, using X-ray and/or SZ measurements \citep{Bulbul2010, Ettori2019}. Since computing
cluster masses in this way for large cluster samples can be prohibitively expensive, ``scaling
relations'' between cluster observables (such as luminosity, temperature, gas mass, or combinations
of observables) and total mass computed from smaller samples can be used to estimate masses for
larger samples to be used for estimating cosmological parameters \citep{Bulbul2019, Bahar2022,
Chiu2022}. 

The scaling relations between X-ray observables and masses are typically computed under the assumptions of hydrostatic equilibrium and sphericity of the clusters \citep{Gianfagna2023}. Needless to say, hydrostatic equilibrium is only satisfied to varying degrees in clusters, with mergers driving non-thermal gas motions \citep[see][for a review]{Pratt2019}. The condition of spherical symmetry is also somewhat violated, through mergers and accretion along cosmic filaments, which produces clusters with triaxial and irregular shapes \citep{Becker2011,Lau2011}. 

In addition to the possible biases introduced by non-thermal pressure and asphericity, there are other potential biases introduced by multiphase gas and projection effects. The first bias comes from the fact that the ICM exhibits a range of temperatures and metallicities, though there are typically only enough counts in a low-exposure observation of a distant cluster to fit all of the emission within a particular projected radius (typically $r_{500c}$ or $r_{200c}$)\footnote{The overdensity radius that defines the region within which the density is 500 or 200 times the critical density of the universe} with a single-temperature and metallicity plasma emission model. These single-component models will inevitably not capture the multiphase structure of the gas, biasing the temperature and/or luminosity estimates \citep{Peterson2003,Kaastra2004,Biffi2012,Frank2013}. The second bias arises from structures that are projected in front of or behind the spherical radius of interest that nevertheless contribute to the observed emission. These structures can be associated with the cluster itself at larger radius along the sight line, or from other clusters, groups, and/or filaments projected along the sight line. 

Additionally, ICM temperature is a key ingredient for cluster mass measurements from X-ray observations \citep[e.g.,][]{Bulbul2010}. Besides calibration differences, the multi-phase nature of the ICM and the structures along the line of sight may yield departing temperature measurements due to varying sensitivity of X-ray telescopes \citep{Schellenberger2015}. It is crucial to disentangle these competing effects with simulations to understand the biases in temperature and mass measurements from X-ray observations. In the context of \textit{eROSITA} \citep{Predehl2021}, launched in 2019 on board the Spectrum-Roentgen-Gamma (SRG) mission \citep{Sunyaev2021}, understanding the interplay between the projection effects, multi-phase nature of ICM, and calibration differences will help with the future cross-calibration work \citep{Liu2022c, Sanders2022, Iljenkarevic2022, Veronica2022, Whelan2022} and hydrostatic mass bias \citep{Scheck2022}.

In this work, we seek to address the impact of both the multi-phase gas and projection biases on the cluster observables of temperature and luminosity using mock observations of galaxy clusters from the smoothed particle hydrodynamics (SPH) Magneticum Pathfinder Simulations\footnote{http://www.magneticum.org} \citep{Biffi2013,Hirschmann2014,Dolag2017,Biffi2022}. Specifically, we model the thermal emission from the hot ICM of the clusters, and pass this through an instrument model for \textit{eROSITA} which includes the effects of the 7 separate telescope modules (TMs), PSF, energy-dependent effective area, spectral response, particle background, and instrument noise. We then fit single-temperature plasma models to the resulting spectra and determine the best-fit temperature and luminosity. We carry this analysis out for three separate samples of the galaxy clusters including increasing amounts of material projected along the sight line, in order to determine the bias on the luminosity and temperature induced by the presence of these structures in the observations. 

The rest of this work is structured as follows: in Section \ref{sec:methods} we describe the
Magneticum simulations and the cluster sample taken from them, as well as the methods used to create
the synthetic X-ray observations of the clusters and fit the resulting spectra to obtain the
relevant observables. In Section \ref{sec:results} we detail the results of our study, and in
Section \ref{sec:conc} we present our conclusions.  

\section{Methods}\label{sec:methods}

\subsection{Simulations and Cluster Sample}\label{sec:methods_sample}

The Magneticum simulations \citep{Hirschmann2014,2015MNRAS.451.4277D} were run using the TreePM/SPH code P-Gadget3, an extended version of
P-Gadget2 \citep{Springel2005}. Beyond hydrodynamics, gravity, and evolution of the collisionless DM
component, the simulations also include radiative cooling and heating from a time-dependent UV
background, star formation and feedback, metal enrichment, and black hole growth and AGN feedback.
More details about the physics implemented in the simulation can be found in \citet{Biffi2022},
their Section 2 and references therein. 

The ``Box2\_hr'' simulation box comprises a co-moving volume of ($352 h^{-1}$ cMpc)$^3$ and is
resolved with 2$\times$1584$^3$ particles, corresponding to mass resolutions for DM of $m_{\rm DM} =
6.9 \times 10^8 h^{-1} M_\odot$ and gas of $m_{\rm gas} = 1.4 \times 10^8 h^{-1} M_\odot$. The
simulations employ a $\Lambda$CDM cosmology with the Hubble parameter $h$ = 0.704, and the density
parameters for baryons, matter, and dark energy are $\Omega_b = 0.0451$, $\Omega_M = 0.272$, and
$\Omega_\Lambda = 0.728$. The normalization of the fluctuation amplitude at 8~Mpc $\sigma_8 = 0.809$
\citep[from the 7-year results of the Wilkinson Microwave Anisotropy Probe,][]{Komatsu2011}.  

\begin{figure*}
\centering
\includegraphics[width=0.95\textwidth]{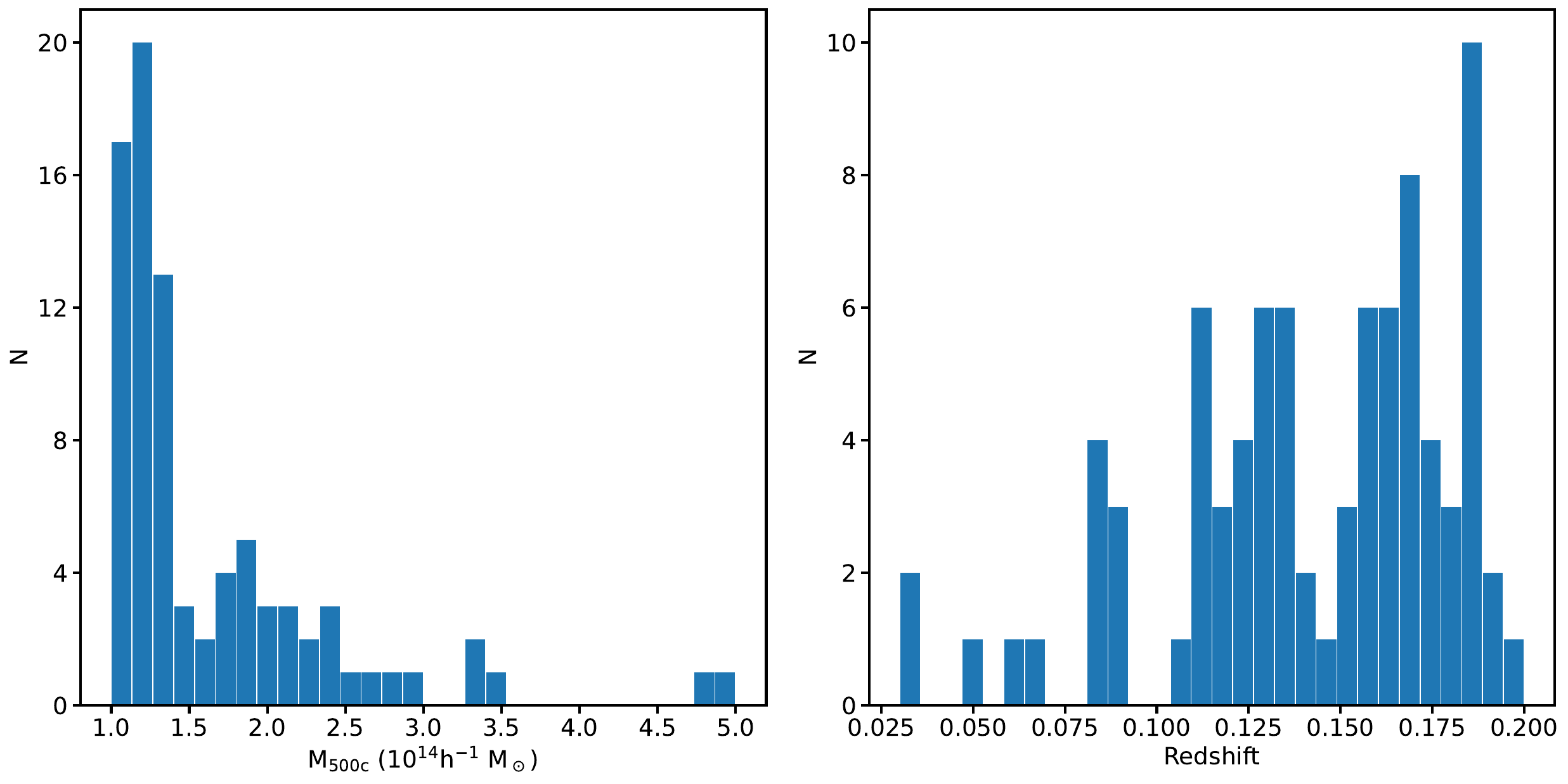}
\caption{Histograms of $M_{500c}$ (left) and the redshift $z_{\rm true}$ for the 84 clusters in the sample.\label{fig:histograms}}
\end{figure*}

Clusters and their substructures were identified using the \code{SubFind} algorithm
\citep{Springel2001,Dolag2009}, which employs a standard friends-of-friends algorithm
\citep{Davis1985}. 84 clusters were selected for this study, using a mass cut of $M_{500c} > 10^{14}
M_\odot/h$, within a lightcone constructed from the simulation which has a FoV of $30 \times 30$ deg$^2$ and a depth of $z$ < 0.2, consisting of 5 independent slices between $z$ = 0.03 and 0.18 (see Table \ref{tab:redshift} for the redshifts of the individual snapshots and the numbers of clusters chosen from each snapshot). Slices have been extracted from each corresponding output box of the Magneticum ``Box2\_hr'' simulation by randomly shifting the pointing direction within the box. The redshift used in computing distances $z_{\rm true}$ is obtained by computing the offset of the cluster center from the center of the slice, and the redshift used in fitting spectra $z_{\rm obs}$ also takes into account the peculiar velocity of the cluster within the slice. Figure \ref{fig:histograms} shows histograms of cluster masses and redshifts from the sample. The center of each cluster is identified as the potential minimum. 

\subsection{Creating Photon Lists with \code{PHOX}}\label{sec:methods_phox}

From our Magneticum cluster sample, we create simulated X-ray photons using the \code{PHOX} software
package \citep{Biffi2012,Biffi2013}. \code{PHOX} takes every gas particle in the simulation and
computes the expected thermal X-ray spectrum from it, based on the particle's density, temperature,
and abundance information. For this work, the spectra have been determined using version 2.0.1 of
the APEC model \citep{Smith2001}. Given the spectrum for each particle, we assume large values for
the exposure time $t_{\rm exp}$ and a ``flat'' effective area $A_{\rm eff}$ to generate a large
sample of X-ray photons at the positions of the gas particles using Poisson sampling. The values of
$t_{\rm exp}$ and $A_{\rm eff}$ are much larger than the values which will be employed for the
mock observations described in Section \ref{sec:methods_sixte}, as the ``observed'' photons
will be drawn from this sample based on the ``true'' $t_{\rm exp}$ and the effective area curve for
the simulated \textit{eROSITA} instrument. These photon positions are projected along a chosen line
of sight, and their energies are Doppler-shifted according to the line-of-sight velocity of their
originating particles. The energies are then cosmologically redshifted and a fraction of the
redshifted photons are absorbed by Galactic neutral hydrogen, assuming a \code{wabs} \cite{}
absorption model and setting the equivalent hydrogen column density parameter to $N_H = 10^{20}$~cm$^{-2}$. The remaining photons are stored in \code{SIMPUT}\footnote{http://hea-www.harvard.edu/heasarc/formats/simput-1.1.0.pdf} photon lists to be used in the instrument simulation (Section \ref{sec:methods_sixte}). 

\begin{table}
\centering
\caption{Redshifts and Cluster Numbers of Simulation Snapshots}
\label{tab:redshift}
\begin{tabular}{ccc}
\hline
snapshot ID & redshift & \# of clusters \\
\hline
124 & 0.174 & 38 \\
128 & 0.137 & 24 \\
132 & 0.101 & 16 \\
136 & 0.066 & 4 \\
140 & 0.033 & 2 \\
\hline
\end{tabular}
\end{table}

We run \code{PHOX} on the lightcone described above. From the slices of the lightcone, we create three separate samples of photon lists. For the ``isolated'' sample, each cluster within the lightcone only has the photons within of 2$r_{\rm 500c}$ of the cluster potential minimum included in the sample. The ``surroundings'' sample includes all of the photons within the redshift slice for each cluster, and thus consists of the emission from the cluster and the structures most nearby to it at the same cosmic epoch in projection. Finally, the ``lightcone'' sample includes the full lightcone of emission in projection, including all structures in projection within the simulated redshift range.

\subsection{Creating Event Files with \code{SIXTE}}\label{sec:methods_sixte}

We generate mock \textit{eROSITA} event files using the ``Simulation of X-ray Telescopes'' (\code{SIXTE}) software package
\citep{Dauser2019}, version 2.7.0. Version 1.8.2 of the \textit{eROSITA} instrument model was used. It is the official end-to-end simulator for \textit{eROSITA} and includes all seven TMs separately. \code{SIXTE} traces the photons through the optics by using the measured PSFs and vignetting curves \citep{Predehl2021} onto the detector. The detection process itself includes a detailed model of the charge cloud and read-out process. Specifically, we use a Gaussian charge cloud model with parameters based on ground calibration measurements \citep[see][for a recent comparison to in-flight data]{Koenig2022}. In \code{SIXTE}, five of these (TMs 1, 2, 3, 4, and 6) have identical effective area curves, and TMs 5 and 7 have identical effective area curves, due to the absence of the aluminum on-chip optical light filter that is present on the other TMs \citep[see Section 9.2 of][]{Predehl2021}. All 7 TMs use the same redistribution matrix file (RMF) and a low-energy threshold of 60\,eV.

Each \code{SIMPUT} photon list (3 lists for 84 clusters) is exposed for 2~ks in pointing mode\footnote{Experiments with the ``toy models'' presented in Section \ref{sec:multi_temps} with the pointing and survey responses show no substantial difference between the two modes}. The
aimpoint for each observation is set to the center of each cluster. No background events were added
for any of our \code{SIXTE} simulations---the method of adding background to the spectra is detailed
in Section \ref{sec:methods_spectra}. 

\subsection{Making and Fitting Spectra}\label{sec:methods_spectra}

From the \code{SIXTE}-produced event files, we use the HEASARC FTOOLS tool \code{extractor} (from \code{HEASOFT} v6.21) to
extract a PI spectrum from each \textit{eROSITA} TM. For each cluster, we extract all photons from
within a circle of projected radius $r_{500c}$ centered on the cluster potential minimum. We co-add
the counts from the 7 TMs into two spectra, one for each of the groups with the same effective area
as noted above. 
 
For the background, we implement two components, the cosmic X-ray background (CXB) and the particle
non-X-ray background (NXB) associated with the detector. For the CXB, we assume the form
\code{apec+wabs*(apec+powerlaw)} and the parameters from \cite{McCammon2002}. For the NXB, we employ
a model comprised of a continuum with a number of emission lines added to it from \citet{Liu2022b}, based on analysis of
early ``filter-wheel-closed'' and the \textit{eROSITA} Final Equatorial Depth Survey (eFEDS) data \citep{Brunner2022, Liu2022a, Bulbul2022}.
Instead of including the background events in each \code{SIXTE} simulation, we instead generate
background PI spectra from the combined CXB+NXB model for a 2~ks exposure and the same extraction region for the source spectra and add them to the cluster spectra. 

For each cluster, we fit the spectra from the two TM groups jointly using \code{XSPEC}, restricting
the fit to the energy range 0.4--7.0~keV, and we use the C-statistic \citep{Cash1979,Kaastra2017}. For the cluster emission, we
use an absorbed thermal model, \code{wabs*apec}. We use APEC v2.0.1 in the fits, as was used in the generation of the photons. We fix the value of the hydrogen column parameter
to $N_H = 10^{20}$~cm$^{-2}$, the same value that was used in the generation of the photon lists. We
fix the metallicity parameter to $Z = 0.3~Z_{\odot}$ in the fits as it is observationally motivated that cluster metallicity averages at that value \citep{Ezer2017, Mernier2022}. The redshift parameter is held
fixed to the redshift of the cluster determined from the lightcone, and the temperature $kT$ and
normalization parameters are left free to vary. For the background, the overall normalizations of
the CXB and NXB are left free to separately vary with a Gaussian prior of 5\% of the model
normalization, but the rest of the background parameters are held fixed. We use the same cosmological parameters as used for the Magneticum simulation and described in Section \ref{sec:methods_sample}. For a comparison of cluster
fits with and without background, see Appendix \ref{sec:bkgnd}. Unless otherwise noted, all quoted
uncertainties are at the 1-$\sigma$ level. 

\section{Results}\label{sec:results}

\begin{figure*}
\centering
\includegraphics[width=0.47\textwidth]{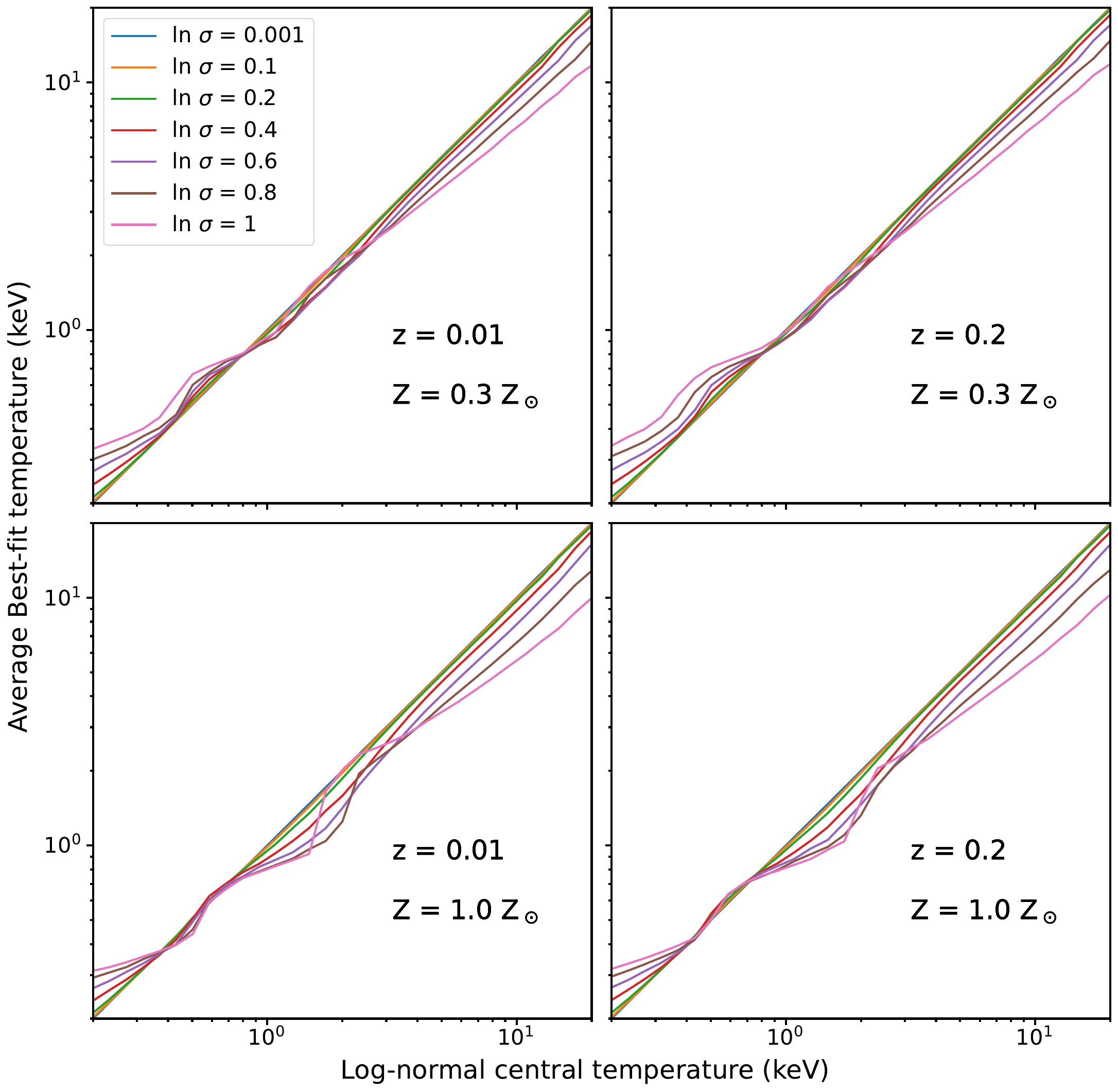}
\includegraphics[width=0.47\textwidth]{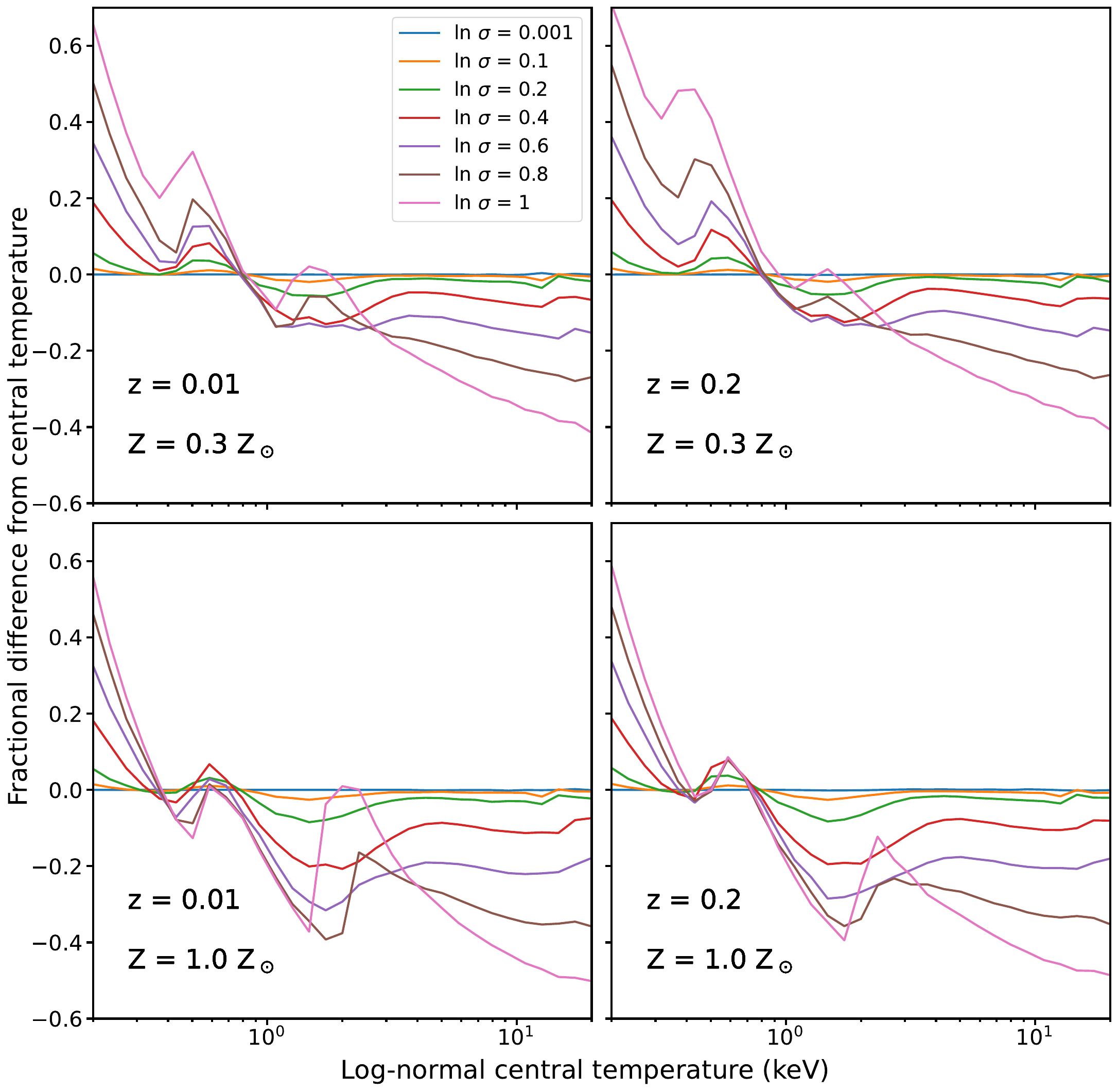}
\caption{Results of a fitting of spectra generated from log-normal temperature distributions, as described in Section \ref{sec:multi_temps}. Left 2x2 panels: central temperature of distribution vs. best-fit temperature. Panels reflect variations in metallicity and redshift. Right 2x2 panels: difference in the best-fit temperature and the central temperature vs. the central temperature for the same distributions.\label{fig:multi_temp}}
\end{figure*}
       
\begin{figure*}
\centering
\includegraphics[width=0.97\textwidth]{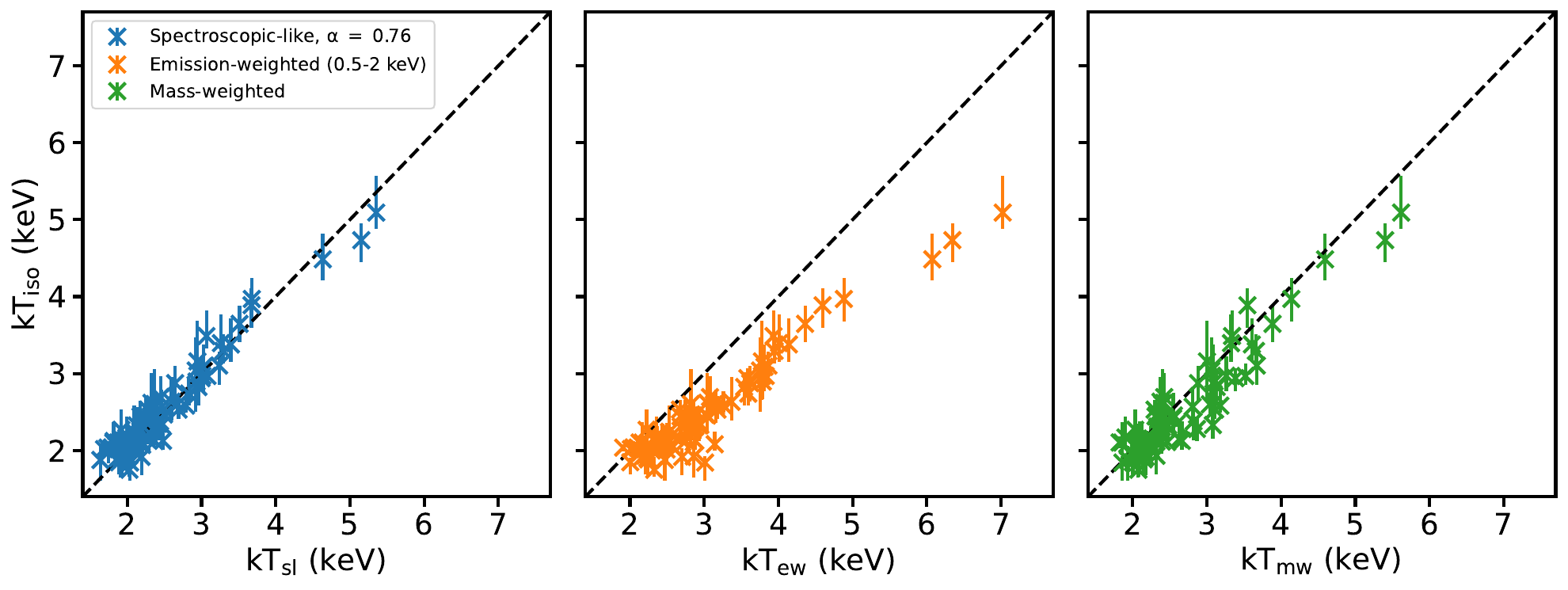}
\caption{Fitted temperatures from the ``isolated'' sample ($kT_{\rm iso}$) plotted against various relevant
weighted temperatures. The black dashed line indicates equality between the two temperatures in each
panel.\label{fig:kT_various}}
\end{figure*}

\begin{figure*}
\centering
\includegraphics[width=0.95\textwidth]{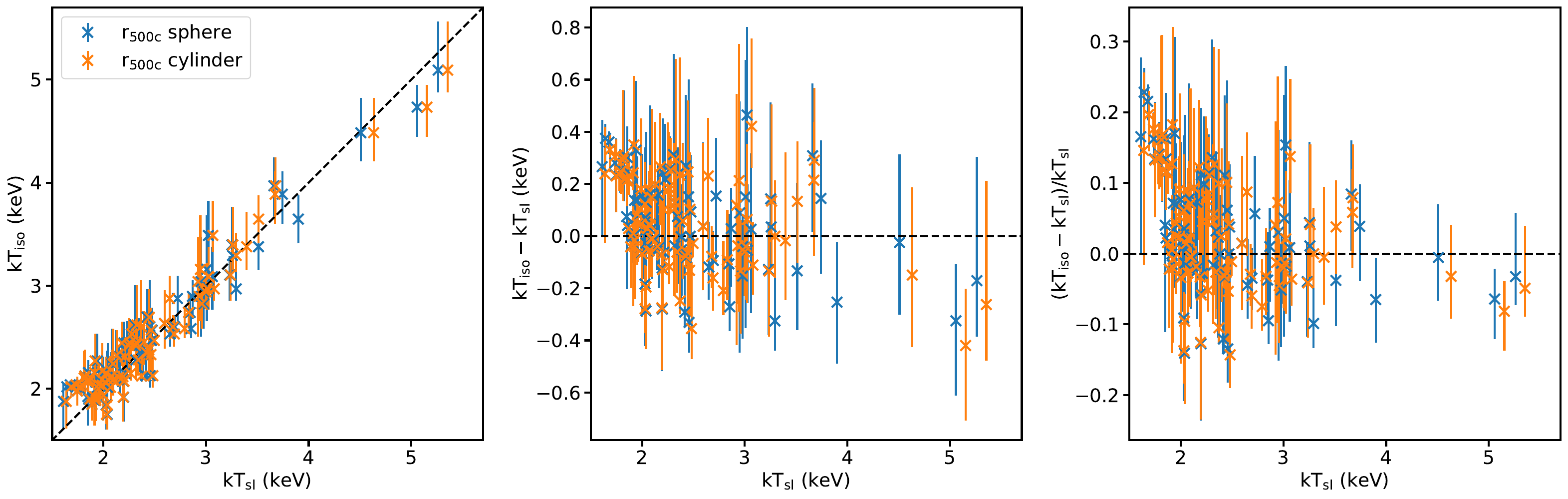}
\caption{The spectroscopic-like temperatures computed from the simulation (using Equation
\ref{eqn:tsl}), using both the sphere and cylinder regions, vs. the fitted temperatures from the
``isolated'' sample. The center panel shows the differences between the two temperatures plotted
against the spectroscopic-like temperatures, and the right panel shows the fractional difference against the spectroscopic-like temperatures. The black dashed lines indicate equality between the temperatures in each panel.\label{fig:Tsim}}
\end{figure*}       

\begin{figure}
\centering
\includegraphics[width=0.48\textwidth]{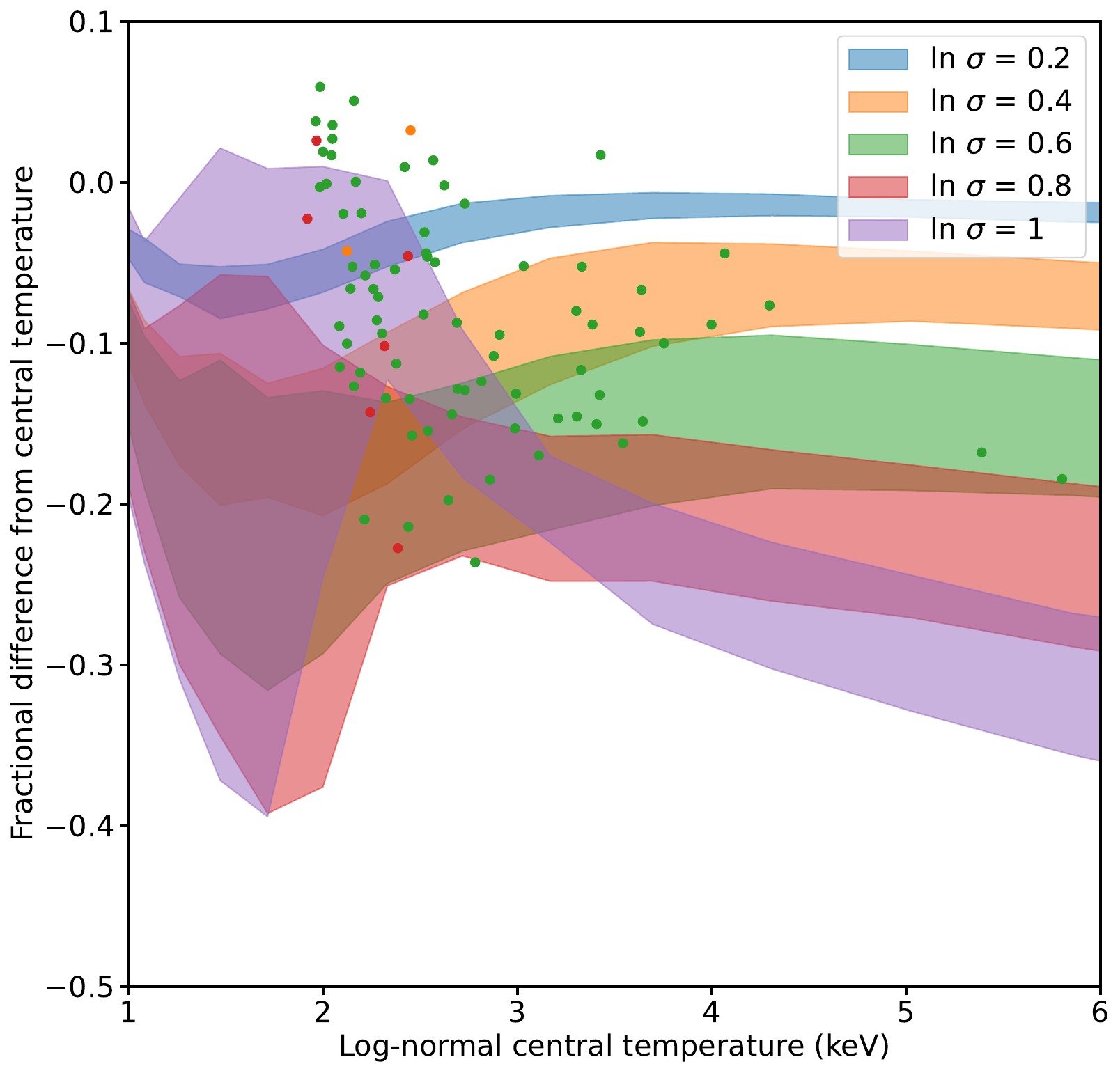}
\caption{Fractional difference of the fitted temperature from the log-normal central temperature $kT_0$ from the toy models and the 84 clusters from the simulation. The bands show the range of values for the four different combinations of redshift and abundance from the toy models described in Section \ref{sec:multi_temps}. The points show the values computed from the clusters, where $\ln{kT_0}$ and $\ln{\sigma}$ have been computed using sample means and variances weighted by the emission measure of the SPH particles. The colors of the points are coded according to the value of $\ln{\sigma}$ they are closest to (within $\Delta{\ln{\sigma}}$ = 0.1).\label{fig:tdiffs}}
\end{figure}
  
\begin{figure*}
\centering
\includegraphics[width=0.95\textwidth]{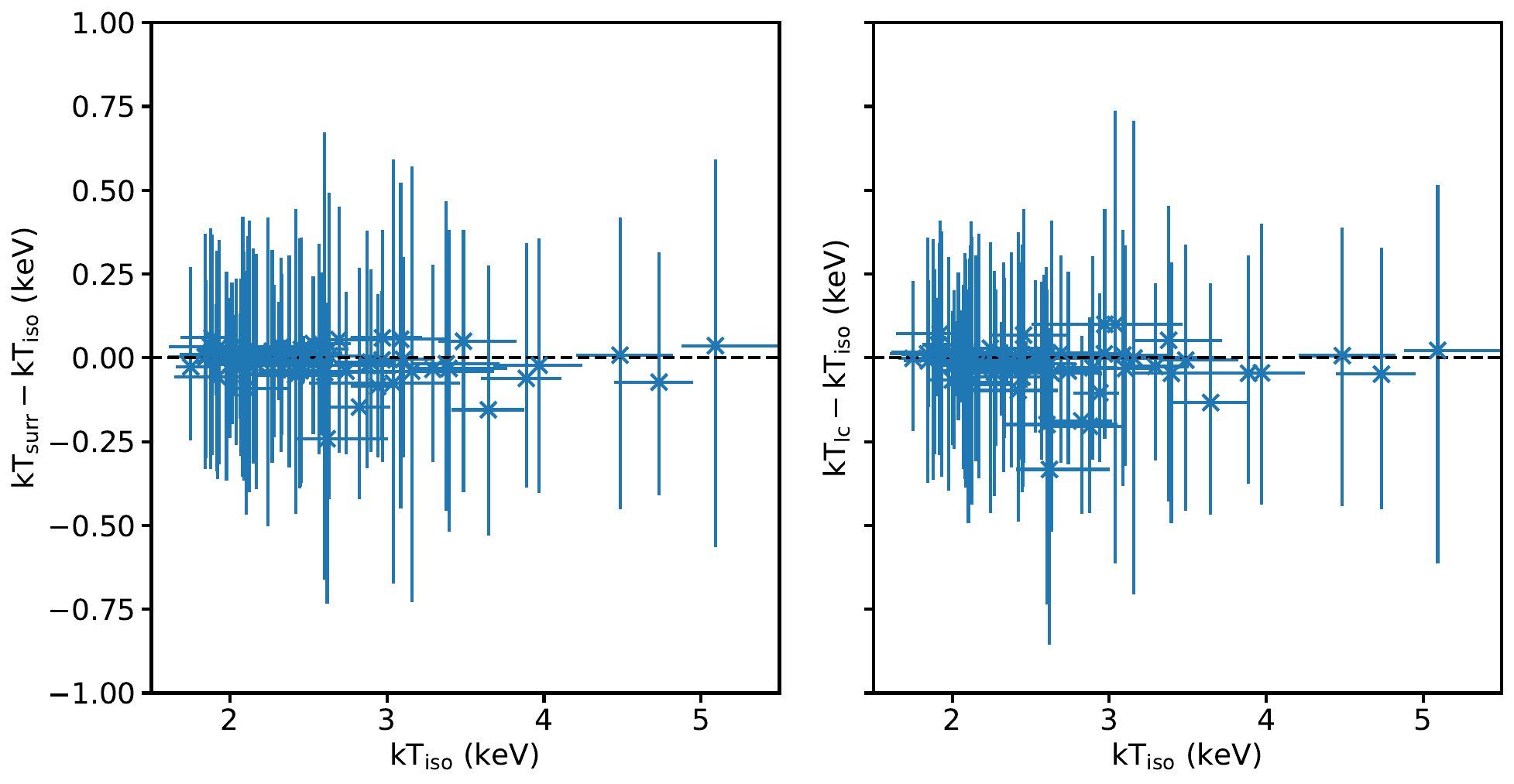}
\caption{Comparisons between the fitted cluster temperatures of the ``isolated'' sample vs. the
``surroundings'' and ``lightcone'' sample, showing the difference of two
samples plotted against the ``isolated'' sample (where the black dashed line indicates no
difference).\label{fig:kTs}}
\end{figure*}
       
\subsection{Toy Models: Effect of Multi-Temperature Structure on Fitted Temperatures}\label{sec:multi_temps}

Before analyzing the temperatures of the clusters in our sample, it is instructive to explore the
effect of multi-temperature structure on the results of single-temperature-model fits using simple
toy models. To this end, we create idealized \code{apec} spectra in \code{XSPEC} with a log-normal
temperature distribution with a central $\ln{kT_0}$ and a $\ln{\sigma}$. We simulate spectra at
redshifts of $z = 0.01$ and $z = 0.2$ (bracketing the bounds of the redshift range of our cluster
sample), and with metallicities of $Z = 0.3$ and $1~Z_\odot$. The mock spectra have foreground
Galactic absorption applied using the \code{phabs} model, for which the column density is $N_H =
10^{20}~{\rm cm}^{-2}$. The spectra are simulated for 40~ks with the \textit{eROSITA} ARF and RMF.
Each spectrum is then fit with a single absorbed \code{phabs*apec} model over the 0.3--7~keV band
assuming no background. This process is repeated 50 times, and an average fitted temperature is
taken from the sample. 

Figure \ref{fig:multi_temp} shows the results of this procedure. The left set of 2x2 panels shows
the recovered fit temperature vs. the input central temperature for a range of $\ln{\sigma}$ for the
different redshift and metallicity options. As should be expected, the fitted temperature is less
well-recovered for larger values of $\ln{\sigma}$. Depending on the values of metallicity and
redshift, the central temperature $kT_0$ is recovered most accurately for values of $\sim
0.5-1.5$~keV. As the spread of temperatures in the distribution increases, the disagreement between
the fitted temperature and the central temperature increases -- it is higher for lower input
temperatures and lower for higher input temperatures. This is due to the decrease in the
\textit{eROSITA} effective area at both low and high energies, making it more difficult to
accurately constrain low and high temperatures. The right set of 2x2 panels shows the difference
between the fitted and the central temperature vs. the central temperature. For $\ln{\sigma} \leq 0.2$, the relative error is less than $\sim$5\% for all temperatures in the range 0.2--20~keV. The error at temperatures $ 0.4 \lesssim kT_0 \lesssim 1$~keV is $\lesssim$20\% for $\ln{\sigma} \leq 0.6$, but is $\sim$30\% for higher values. For lower temperatures and higher values of $\ln{\sigma}$, the fractional error can increase to $\sim$50--60\%. At temperatures $kT_0 \gtrsim 1$~keV, the fractional errors increase with increasing temperature, down to a $\sim$20\% decrease for $\ln{\sigma} \leq 0.6$, and down to a $\sim$50\% decrease for $\ln{\sigma} \leq 1$. In general, the errors at higher temperatures ($kT_0 \gtrsim 1$~keV) are larger for higher redshifts and higher metallicities. 

\subsection{Cluster Temperatures}

\subsubsection{Comparison of Fitted Temperatures with Simulation Temperatures}\label{sec:sim_temps}

We begin the analysis of the cluster mocks by comparing the fitted spectral temperatures from the mock \textit{eROSITA} observations
to cluster temperatures determined from the simulation data with different weightings. For this, we
use the fitted temperatures $kT_{\rm iso}$ from the ``isolated'' sample to determine a
``spectroscopic-like'' temperature, employing the method of \citet{Mazzotta2004} (hereafter M04).
The general idea is that the temperature estimated via spectral fitting can be approximated by a
weighting of the form
\begin{equation}\label{eqn:tsl}
T_{\rm sl} = \frac{\displaystyle\int{wTdV}}{\displaystyle\int{w~dV}}.
\end{equation}
where the weighting function is:\footnote{Note that in M04 the weighting is $\propto
T^{\alpha-3/2}$, whereas in our case we absorb the $-3/2$ into the definition of $\alpha$, and
change the sign of the exponent so that $\alpha > 0$.}  
\begin{equation}\label{eqn:weight}
w = n_en_pT^{-\alpha},
\end{equation}
where $T$ is the temperature of each SPH particle, and $n_e$ and $n_p$ are the electron and proton
densities, respectively. For {\it Chandra} and {\it XMM-Newton} observations, M04 found a best-fit
value of $\alpha \approx 0.75$. Since the effective area of \textit{eROSITA} is different from {\it
Chandra} and {\it XMM-Newton}, we cannot simply assume the same value of $\alpha$, but instead must
determine it via a similar procedure. 

To do this, we first compute $T_{\rm sl}$ from Equation \ref{eqn:tsl} for each of the clusters in
our sample within a cylinder of radius of $r_{500c}$, centered on the cluster potential minimum and extended along the line of sight (using only the gas particles as belonging to the cluster as identified by the \code{SUBFIND} algorithm), for a range of $\alpha$ values. We use a cylinder instead of a sphere with radius $r_{500c}$ since the spectrum which is fit includes emission from along the entire line of sight within the projected angular radius corresponding to $r_{500c}$. Then, following a similar procedure to M04, we determine the relative error in $T_{\rm sl}$, summed over all
of the clusters:
\begin{equation}
\Delta(\alpha) = \frac{1}{N}\sqrt{{\sum_i}\left(\frac{T_{{\rm sl},i}-T_{{\rm iso},i}}{T_{{\rm iso},i}}\right)^2}
\end{equation}
We then minimize $\Delta(\alpha)$ to find a best-fit value of $\alpha = 0.76 \substack{+0.05 \\
-0.04}$. Error bars on $\alpha$ were determined by re-sampling 1000 different values of the fitted
temperature for each cluster in the ``isolated'' sample, fitting for $\alpha$ for each of the
1000 samples, and finding the 68\% confidence limit. This value is consistent within the 1-$\sigma$ errors to that found for \textit{Chandra} and \textit{XMM-Newton} temperatures from M04. 

Figure \ref{fig:kT_various} shows $kT_{\rm iso}$ plotted against the spectroscopic-like temperature $kT_{\rm sl}$, the emission-weighted temperature $kT_{\rm ew}$ in the 0.5--2~keV band, and the mass-weighted temperature $kT_{\rm mw}$. Each temperature measure is computed for each cluster directly from the simulation data. The leftmost panel shows
$kT_{\rm iso}$ versus $kT_{\rm sl}$ with $\alpha = 0.76$. Despite
the simplicity of this formula, the relation represents reasonably well the distribution of fitted
temperatures with a mean difference between $kT_{\rm iso}$ and $kT_{\rm sl}$ of $\sim$0.05~keV
($\sim$3\%) and a standard deviation of 0.18~keV ($\sim$8\%, see also Table \ref{tab:sim_stats}).
$kT_{\rm ew}$ (center panel) trends higher than the fitted
spectroscopic temperature, especially at temperatures emitting strongest in rest-frame energies $\gtrsim 2$~keV here the sensitivity of \textit{eROSITA} decreases. Even though the 0.5--2~keV band is covered well by \textit{eROSITA}'s effective area, the clusters at higher temperatures which only
contribute photons at the higher end of this band are downweighted in the best-fit temperatures
compared to the emission-weighted temperatures. The same underestimate in best-fit temperature is
also seen in the comparison to $kT_{\rm mw}$ (right panel), though not as severely.

\begin{table*}
\centering
\caption{Mean and Standard Deviation in Differences Between Fitted Quantities and Simulations}
\label{tab:sim_stats}
\begin{tabular}{ccc}
\hline
& $kT_{\rm iso}-kT_{\rm sl}$ sphere/cylinder & $(L_{\rm iso}-L_{\rm sim})E(z)^{-1}$ sphere/cylinder \\
& (keV) & ($10^{44}$~erg~s$^{-1}$) \\
\hline
mean & 0.05/0.05 & 0.030/-0.006 \\
std. dev. & 0.18/0.18 & 0.036/0.032 \\
mean \% & 2.7/2.9 & 6.3/-2.3 \\
std. dev. \% & 7.9/7.8 & 4.6/5.1 \\
\hline
\end{tabular}
\end{table*}

Figure \ref{fig:Tsim} shows $kT_{\rm iso}$ plotted against $kT_{\rm sl}$ again (left panel), along
with the difference between the two temperatures $kT_{\rm iso}-kT_{\rm sl}$ plotted against $kT_{\rm
sl}$ (center panel), and the fractional difference $(kT_{\rm iso}-kT_{\rm sl})/kT_{\rm sl}$. In this
figure, we also show the effect of computing $kT_{\rm sl}$ using all of the gas particles within a
sphere of radius $r_{500c}$ centered on the cluster potential minimum (blue points) in addition to
the temperatures computed from the cylindrical regions. The latter is more representative of what is
measured by the spectral fitting, since everything is included in projection within the aperature of
$r_{500c}$. For both cases, most of the differences between the two temperatures fall between $\pm$0.4~keV, or $\pm$20\%. The overall distributions of the temperatures measured within the
spheres or the cylinders are very similar (Figure \ref{fig:Tsim} and Table \ref{tab:sim_stats}),
though the power-law index in Equation \ref{eqn:weight} for the spherical case is $\alpha = 0.85
\substack{+0.05 \\ -0.03}$, roughly 2-$\sigma$ away from the value of $\alpha = 0.76$ computed for
the cylindrical regions. The overall similarity between the two temperatures reflects the fact that it is dominated by the gas with higher emission measures ($\propto n^2$) near the cluster centers, which will be similarly captured by either the cylindrical or spherical regions. 

It can be seen from the center panel of Figure \ref{fig:Tsim} that the errors in $kT_{\rm sl}$ skew
slightly towards lower best-fit temperatures at higher $kT_{\rm sl}$. This same trend in all four
temperature measures from Figures \ref{fig:multi_temp} and \ref{fig:kT_various} is consistent with
the results of Section \ref{sec:multi_temps}. Table \ref{tab:sim_stats} shows the mean and standard
deviation of $kT_{\rm iso}-kT_{\rm sl}$, which are relatively small. However, M04 advised that for
\textit{Chandra} and \textit{XMM-Newton} spectra the simple formula for $kT_{\rm sl}$ is accurate to
a few to several percent only for temperatures of $kT \gtrsim 3$~keV, where the spectrum is
dominated by continuum emission. \citet{Vikhlinin2006} showed that for $kT \lesssim 3$~keV a more
complicated (and non-analytic) algorithm for determining $kT_{\rm sl}$ was required for these
line-dominated spectra. Most of our sample lies in this lower-temperature range, which is compounded
by the fact that \textit{eROSITA} is more sensitive at lower X-ray energies, ensuring that our
cluster spectra are mostly dominated by line emission from metals. For this reason, the simple
power-law prescription in our case is less accurate, with a standard deviation of $\sim$8\% (Table
\ref{tab:sim_stats}) and maximum deviations of $\sim \pm$20\% (right-most panel of Figure
\ref{fig:Tsim}), especially at lower temperatures which are the most line-dominated. We save a
treatment similar to \citet{Vikhlinin2006} for \textit{eROSITA} spectra for future work.

We can also compare the fitted temperatures $kT_{\rm iso}$ and their differences from the expected temperature from the simulation to the results of the toy models in Section \ref{sec:multi_temps}. To do this, we compute the sample mean $\ln{kT_0}$ and the sample standard deviation $\ln{\sigma}$ from the SPH particles in each cluster, weighted by the emission measure. The former allows us to compare to the fractional difference of the fitted temperature from the central temperature from the toy models directly. This is shown in Figure \ref{fig:tdiffs}. The bands show the range of fractional differences from all four combinations of metallicity and redshift from Figure \ref{fig:multi_temp}, for several values of $\ln{\sigma}$. The points show the values of the fractional difference computed for the 84 clusters, where the colors of the points are coded according to the value of $\ln{\sigma}$ they are closest to (within $\Delta{\ln{\sigma}}$ = 0.1). 

There is not a precise correspondence between the values from the clusters and the toy models--nor should one be expected, since the cluster gas temperatures do not necessarily follow a log-normal distribution. However, there is at least qualitative agreement between them. For most of the 84 clusters, $\ln{\sigma} \approx 0.6$ (green points). For most of the clusters, the fitted temperature $kT_{\rm iso}$ underestimates the central temperature $kT_0$ by $\sim$10-20\%, in general agreement with the predictions of the toy models. 

\subsubsection{The Effect of Cosmic Structure on the Observed Cluster Temperatures}\label{sec:obs_temps}

Other structures aligned with our observed clusters in projection will bias the observed temperature
of the cluster. Figure \ref{fig:kTs} shows the differences between the ``isolated'' sample and the
``surroundings'' and ``lightcone'' samples, plotted against the temperatures from the ``isolated''
sample. The temperatures of the ``surroundings'' and ``lightcone'' samples in general track the
``isolated'' sample very closely, all well within the measurement errors. Table
\ref{tab:sample_stats} shows the mean and standard deviation of $\Delta{kT}$ for each of the two
comparisons shown in Figure \ref{fig:kTs}. The mean difference in all three samples is very small,
with an absolute value $\leq$~0.02~keV in all cases. The standard deviation of the differences
between the samples is also very small, $\sigma_{\rm kT} \approx 0.04-0.06$~keV.

\begin{table*}
\centering
\caption{Mean and Standard Deviation in Changes to Observed Temperature and Luminosity Between Samples}
\label{tab:sample_stats}
\begin{tabular}{ccccc}
\hline
& $kT_{\rm surr}-kT_{\rm iso}$ & $kT_{\rm lc}-kT_{\rm iso}$ & $(L_{\rm surr}-L_{\rm iso})E(z)^{-1}$ & $(L_{\rm lc}-L_{\rm iso})E(z)^{-1}$ \\
& (keV) & (keV) & ($10^{44}$~erg~s$^{-1}$) & ($10^{44}$~erg~s$^{-1}$) \\
\hline
mean & -0.011 & -0.019 & 0.0026 & 0.0064 \\
std. dev. & 0.045 & 0.062 & 0.0079 & 0.0095 \\
mean \% & -0.4 & -0.7 & 0.7 & 1.7 \\
std. dev. \% & 1.7 & 2.3 & 2.4 & 3.0 \\
\hline
\end{tabular}
\end{table*}

\subsection{Cluster Luminosities}

\begin{figure*}
\centering
\includegraphics[width=0.95\textwidth]{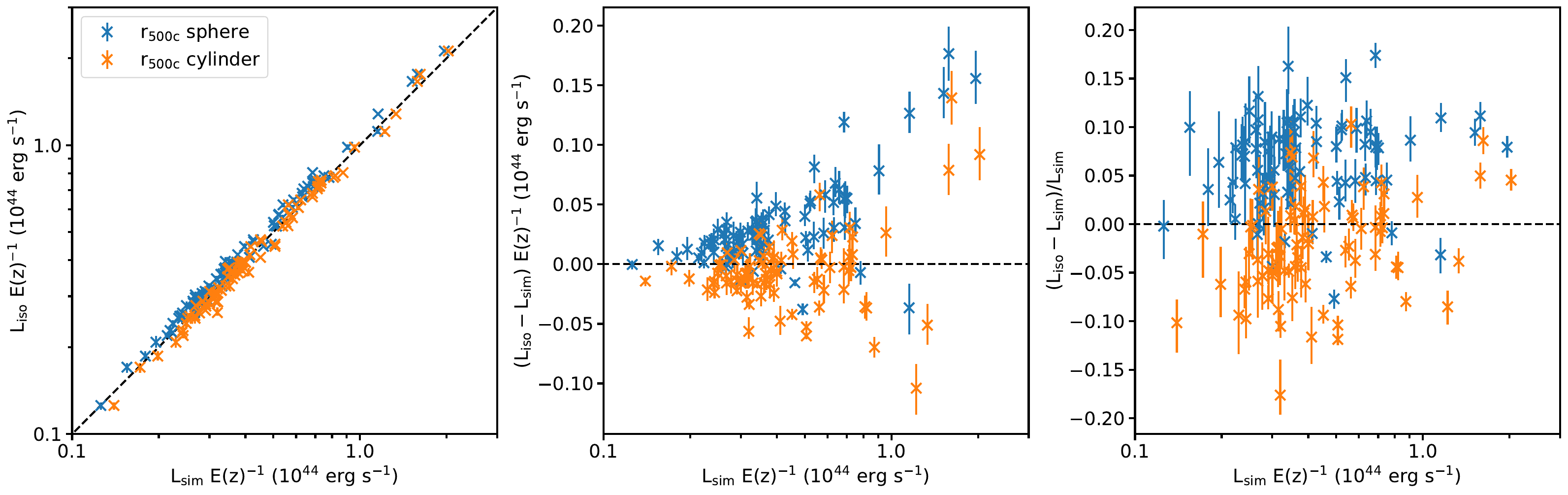}
\caption{Cluster luminosities computed from the simulation vs. the fitted luminosities from the
``isolated'' sample. The right panel shows the differences between the two luminosities plotted
against the simulation luminosity. The black dashed lines indicate equality between the luminosities in each panel.\label{fig:Lsim}}
\end{figure*}
        
\begin{figure*}
\centering
\includegraphics[width=0.95\textwidth]{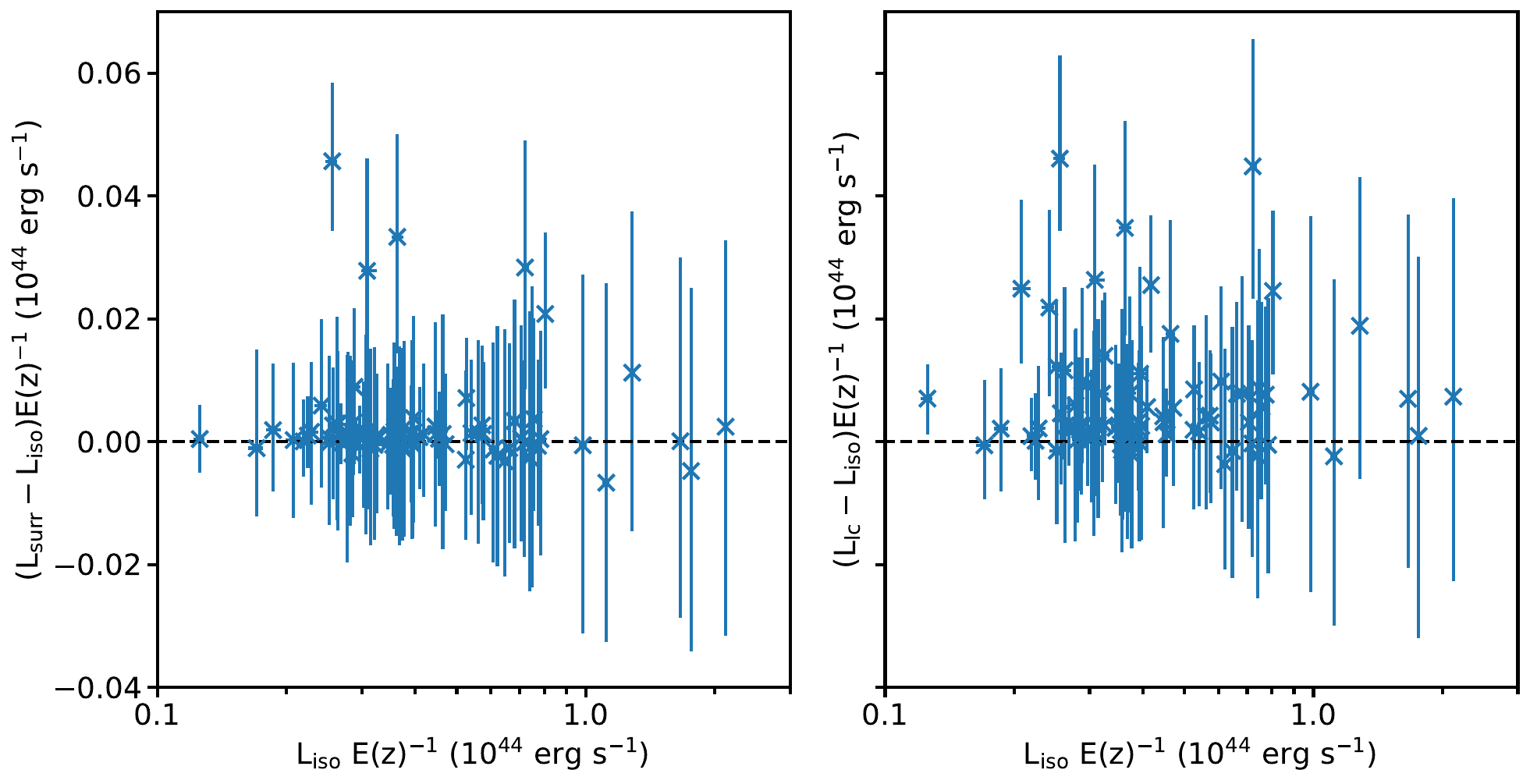}
\caption{Comparisons between the fitted cluster luminosities of the ``isolated'' sample vs. the
``surroundings'' and ``lightcone'' samples, showing the difference of two
samples plotted against the ``isolated'' sample (where the black dashed line indicates no
difference)\label{fig:lums}}
\end{figure*}
  
\subsubsection{Comparison of Fitted Luminosities with Simulation Luminosities}\label{sec:sim_lums}

The luminosities of the clusters from the simulation can be directly compared to the luminosities
estimated from the spectral fitting. The left panel of Figure \ref{fig:Lsim} shows the X-ray
luminosity in the 0.5--2.0~keV band determined from the best-fit model for each cluster in the
``isolated'' sample vs. the computed luminosity in the same band from the SPH particles within a
sphere of radius of $r_{500c}$. The luminosity is computed in \texttt{PHOX} using the same methods
used to compute the spectrum for each SPH particle as described in Section \ref{sec:methods_phox},
except without including the effects of Poisson statistics. The right panel of the same figure shows
the difference between the fitted luminosity and the simulation luminosity $L_{\rm iso}-L_{\rm
sim}$, for both the sphere and cylinder regions mentioned above. Here, unlike the temperatures in
Section \ref{sec:sim_temps}, the difference between the fitted luminosity and the simulation
luminosity depends very clearly on the region chosen. The fitted luminosity overestimates the
luminosity in the spherical regions by $\sim$6\%, due to the fact that the former includes
material outside of the spherical radius of $r_{500c}$ along the sight line. The fitted luminosity
is in much better agreement with the luminosity in the cylindrical regions, as expected, with a mean
difference of $\sim-2$\%. In both cases, the standard deviation of the luminosity differences is
$\sim$5\%. A similar luminosity bias from a projected measurement over that expected from a spherical region was noted by \citet{Dolag2006}, who also found a similar overestimate of less than $\sim$10\% (see the discussion in their Section 5.2). 

Aside from this bias in the luminosity related to geometrical effects, another fundamental limitation is that of fitting a single-component temperature and abundance model to a plasma which is inherently multi-temperature and with varying chemical composition. The best-fit single-component model will necessarily only capture a portion of the expected luminosity, depending on how much the plasma differs from a single phase. Also, as mentioned above, in the fits the metallicity parameter is held fixed at $Z = 0.3Z_\odot$, which is a typical value outside of the core region of clusters. The typical number of counts in a 2~ks spectrum for any of our clusters do not provide sufficient statistics to constrain the metallicity. The metallicity in the cores of clusters is typically higher, which could lead to an underestimate in the luminosity. 

\subsubsection{The Effect of Cosmic Structure on the Observed Cluster Luminosities}\label{sec:obs_lums}

Substructures in projection with observed clusters will bias the estimated luminosity upward.
Therefore, it should be expected that the ``surroundings'' and ``lightcone'' samples can have
higher luminosities than the ``isolated'' sample. 

Figure \ref{fig:lums} shows comparisons between the luminosities of the clusters from the
``isolated'' sample vs. the ``surroundings'' and ``lightcone'' samples in terms of the difference
${\Delta}L$ between the samples on the $y$-axis. Most of the differences between the ``isolated''
sample and the ``surroundings'' and ``lightcone'' samples are very minor, but there are several
clusters for which the increase in luminosity due to projected structures is somewhat significant
($\sim$5--20\%). The number of these clusters with significant deviations is larger in the
``lightcone'' sample, as expected. Overall, however, the mean difference is very small, with very
low scatter ($\sim$2--3\%), as seen in Table \ref{tab:sample_stats}.

\begin{figure}
\centering
\includegraphics[width=0.49\textwidth]{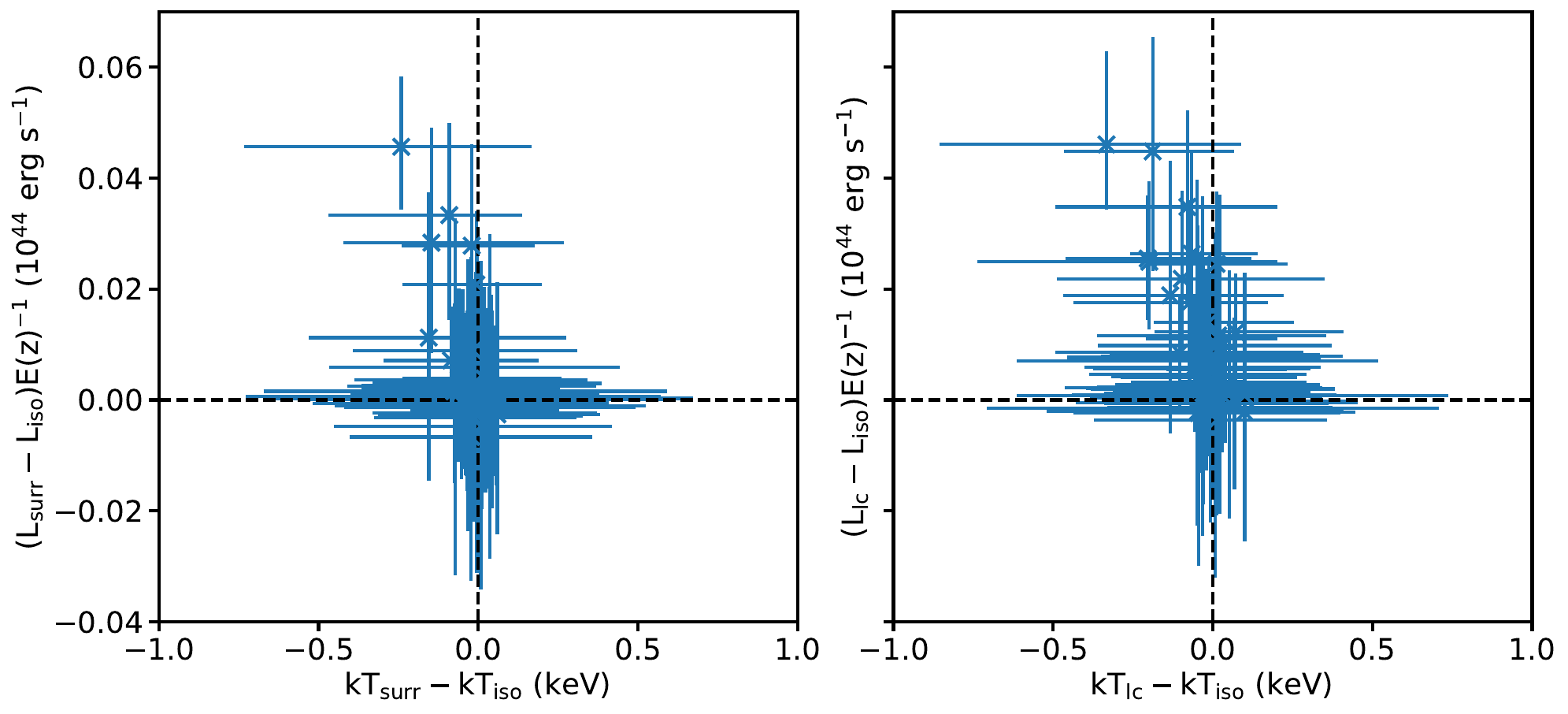}
\caption{Comparisons of differences between the ``isolated'' sample and the ``surroundings'' and ``lightcone'' samples in temperature versus luminosity of the
three different samples. Black dashed lines indicate no difference between the two samples.\label{fig:dL_vs_dT}}
\end{figure}

It is also instructive to examine the differences in temperature and luminosity between the samples together. This is shown in Figure \ref{fig:dL_vs_dT}, which plots the differences in luminosity versus temperature between the ``isolated'' sample and the other two. As already seen, most clusters lie very close to the point of no significant difference in either temperature or luminosity together, but there is a trend of a small subset of clusters with higher luminosity and lower temperature (going up and to the left in both panels of Figure \ref{fig:dL_vs_dT}), which is more statistically significant in luminosity than temperature. This effect is more pronounced in the difference between ``lightcone'' and ``isolated'' samples. The overall effect is readily attributed to the fact that the densest gas in clusters in general is that which has cooled in the cores, so that any bright substructure in projection that makes a significant increase in apparent brightness is also likely to make the target cluster appear cooler than it actually is. 

\subsection{Largest Differences Between the Samples}
 
We will now look at the clusters which have the largest differences in fitted temperature and luminosity between the samples from Sections \ref{sec:obs_temps} and \ref{sec:obs_lums} and visually inspect them. Note that here we also inspect differences between the ``surroundings'' and ``lightcone'' samples, unlike the previous sections. The five clusters we show are:

\begin{itemize}

\item \textbf{snapshot 128, halo ID 231} has the highest $(kT_{\rm surr}-kT_{\rm iso})/kT_{\rm iso}$ = 3.2\% at $kT_{\rm iso}$ = 1.89~keV
\item \textbf{snapshot 128, halo ID 241} has the highest $(kT_{\rm lc}-kT_{\rm surr})/kT_{\rm surr}$ = 6.9\% at $kT_{\rm surr}$ = 1.87~keV
\item \textbf{snapshot 128, halo ID 46} has the lowest $(kT_{\rm surr}-kT_{\rm iso})/kT_{\rm iso}$ = -9.2\% at $kT_{\rm iso}$ = 2.62~keV and the highest $(L_{\rm surr}-L_{\rm iso})/L_{\rm iso}$ = 17.8\% at $L_{\rm iso}E(z)^{-1}$ = $0.256 \times 10^{44}$~erg~s$^{-1}$
\item \textbf{snapshot 132, halo ID 91} has the lowest $(kT_{\rm lc}-kT_{\rm surr})/kT_{\rm surr}$ = -6.4\% at $kT_{\rm surr}$ = 2.85~keV
\item \textbf{snapshot 124, halo ID 135} has the highest $(L_{\rm lc}-L_{\rm surr})/L_{\rm surr}$ = 11.8\% at $L_{\rm surr}E(z)^{-1}$ = $0.208 \times 10^{44}$~erg~s$^{-1}$

\end{itemize}

We do not show the largest negative luminosity differences between the samples since these are very small, which is expected since we do not expect the addition of substructure to bias the luminosity lower. 

\begin{figure*}
\centering
\includegraphics[width=0.95\textwidth]{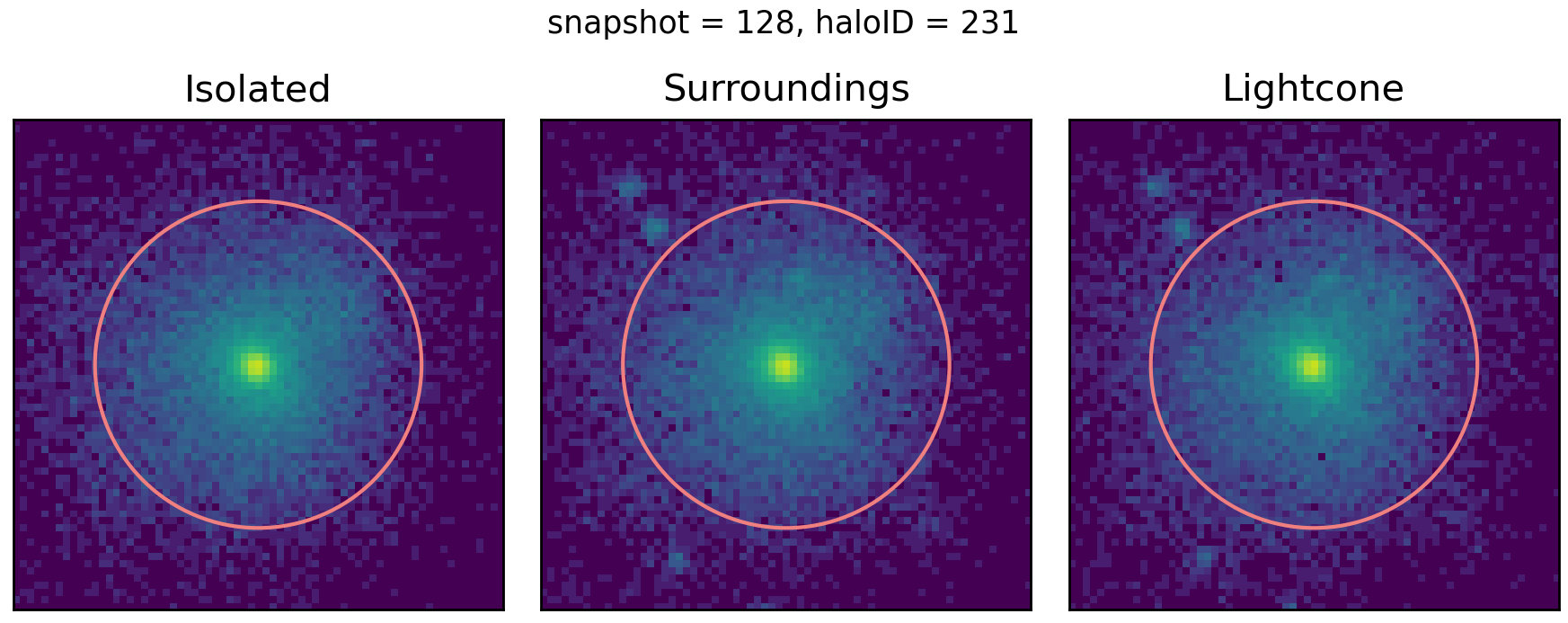}
\caption{Mock \textit{eROSITA} image of the cluster in snapshot 128 with halo ID 231, which has the highest $(kT_{\rm surr}-kT_{\rm iso})/kT_{\rm iso}$. All events are shown and no background is included in the image. The circle indicates a radius of $r_{500c}$.\label{fig:high_T_surr_iso}}
\end{figure*}

\begin{figure*}
\centering
\includegraphics[width=0.95\textwidth]{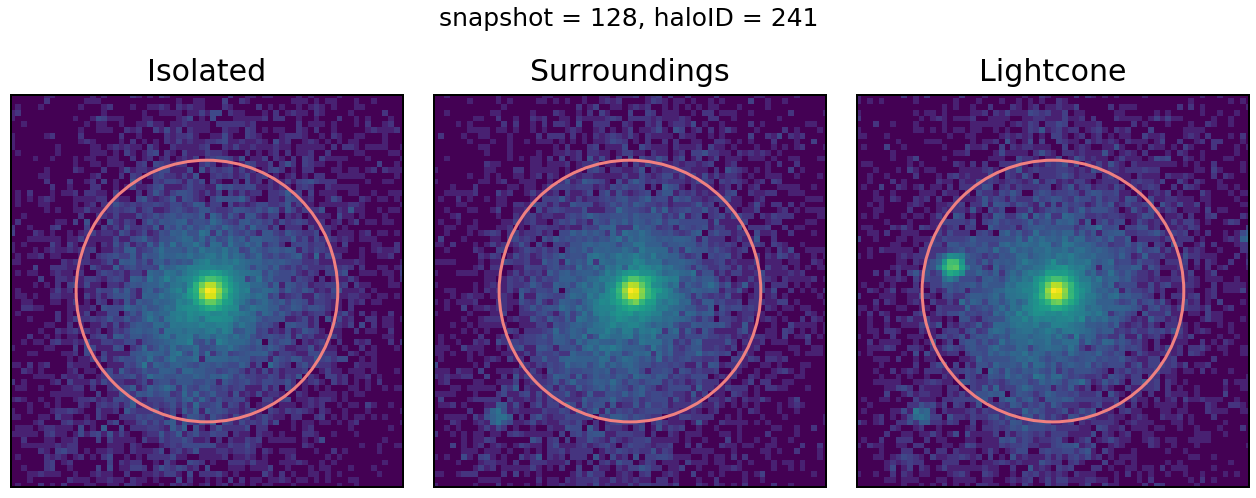}
\caption{Mock \textit{eROSITA} image of the cluster in snapshot 128 with halo ID 241, which has the highest $(kT_{\rm lc}-kT_{\rm surr})/kT_{\rm surr}$. All events are shown and no background is included in the image. The circle indicates a radius of $r_{500c}$.\label{fig:high_T_lc_surr}}
\end{figure*}

\begin{figure*}
\centering
\includegraphics[width=0.95\textwidth]{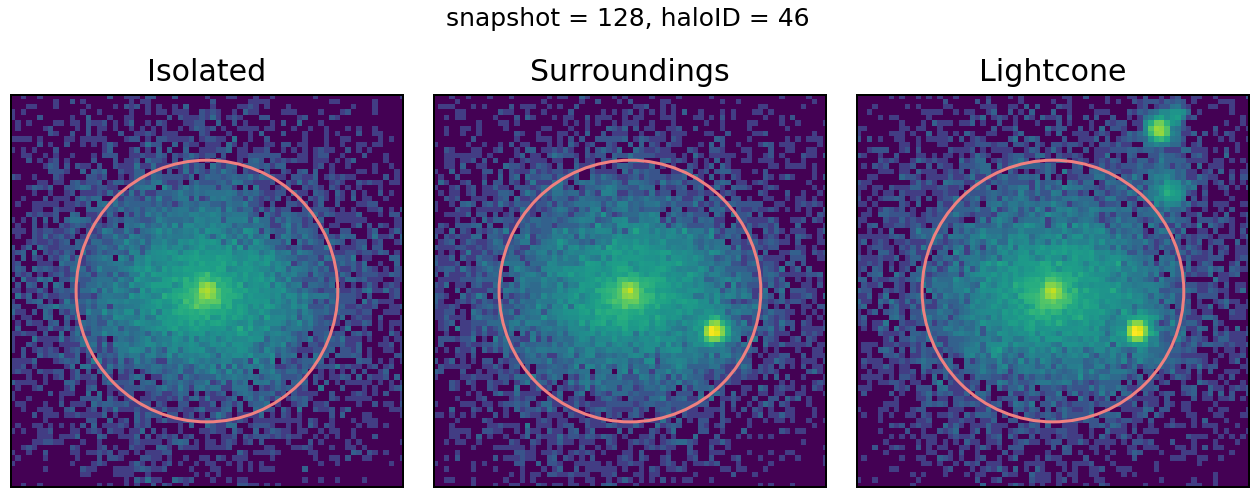}
\caption{Mock \textit{eROSITA} image of the cluster in snapshot 128 with halo ID 46, which has the lowest $(kT_{\rm surr}-kT_{\rm iso})/kT_{\rm iso}$ and the highest $(L_{\rm surr}-L_{\rm iso})/L_{\rm iso}$. All events are shown and no background is included in the image. The circle indicates a radius of $r_{500c}$.\label{fig:low_T_high_L_surr_iso}}
\end{figure*}

\begin{figure*}
\centering
\includegraphics[width=0.95\textwidth]{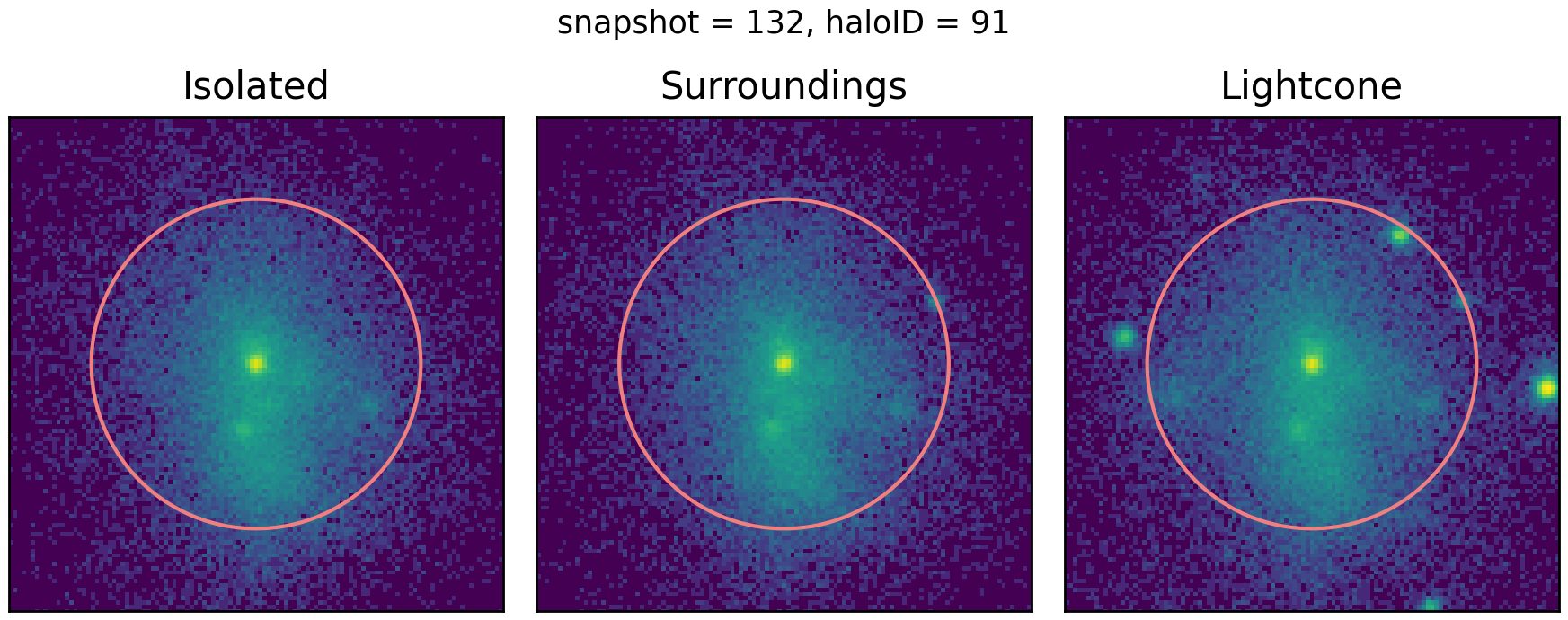}
\caption{Mock \textit{eROSITA} image of the cluster in snapshot 132 with halo ID 91, which has the lowest $(kT_{\rm lc}-kT_{\rm surr})/kT_{\rm surr}$. All events are shown and no background is included in the image. The circle indicates a radius of $r_{500c}$.\label{fig:low_T_lc_surr}}
\end{figure*}

\begin{figure*}
\centering
\includegraphics[width=0.95\textwidth]{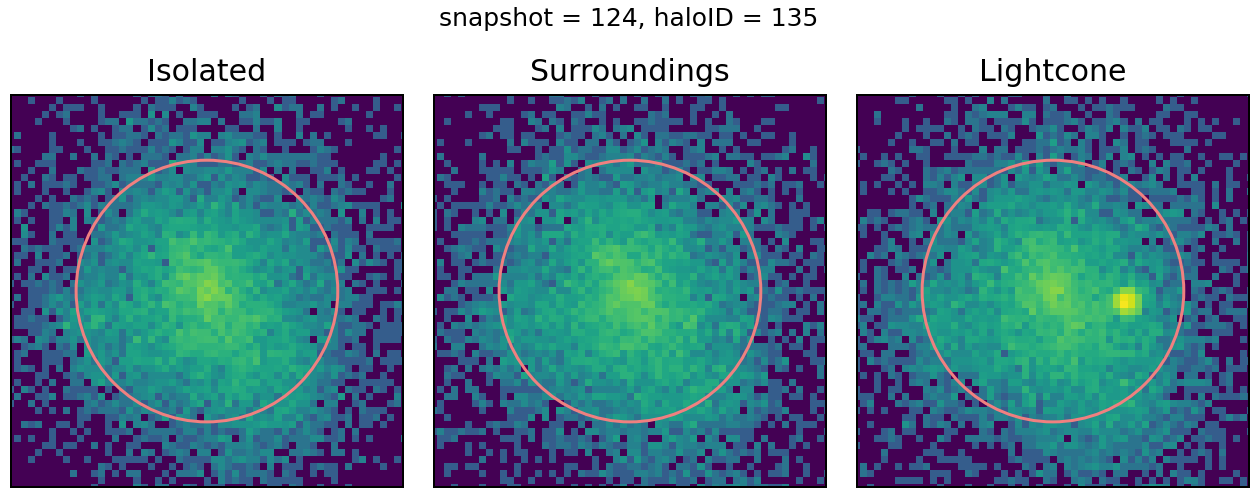}
\caption{Mock \textit{eROSITA} image of the cluster in snapshot 124 with halo ID 135, which has the highest $(L_{\rm lc}-L_{\rm surr})/L_{\rm surr}$. All events are shown and no background is included in the image. The circle indicates a radius of $r_{500c}$.\label{fig:high_L_lc_surr}}
\end{figure*}

We show the mock cluster images (without background) in Figures \ref{fig:high_T_surr_iso}-\ref{fig:high_L_lc_surr}. All detected events are shown. In all of these cases, there is obvious bright substructure that appears within or nearby the aperture of $r_{500c}$ that biases the temperature and/or luminosity. In many cases, such bright substructures may be easily masked to avoid such a luminosity or temperature bias. However, for all of these extreme cases the differences are very small, and for the temperatures are all within the measurement errors. 

\subsection{Luminosity-Temperature Relation}\label{sec:lt_relation}

\subsubsection{L-T Relation: Introduction and Methodology}

The interconnection of the physical properties of clusters is described by scaling relations. \citet{Kaiser1986} derived simple forms of these relations, which are called self-similar scaling relations, by assuming gravity to be dominant during the formation and evolution of clusters. However, it is non-trivial to derive precise forms of these relations because, gravity is not the only dominant process that is regulating the formation and evolution of these objects but there are other complex baryonic processes, such as AGN feedback, that alter the physical properties and therefore the scaling relations \citep{Puchwein2008}. This alteration naturally gives birth to the need for the calibration of these relations using simulations and observations. There is a plethora of studies in the literature that shows the employed sample selection method and criteria may introduce a large bias to the calibrated scaling relations if the selection effects are not properly taken into account \citep[see ][ for examples and discussion]{Mantz2019}. Therefore modeling and calibrating scaling relations go hand to hand with the modeling of the selection and the abundance of the objects as a function of the physical properties of interest.

One of the most affected relations from the non-thermal baryonic processes is the X-ray luminosity-temperature ($L_{\rm X}-T$) relation. This is due to the fact that these observables have a strong dependence on the distribution and the average kinetic energy of the hot ICM that are modified directly by the non-thermal processes. These observables can also be affected by the presence (or absence) of cool cores \citep{Mantz2018,Maughan2007}. $L_{\rm X}$ and $T$ are the two main X-ray observables, therefore there are large number of observational studies trying to constrain the $L_{\rm X}-T$ relation \citep[e.g.,][]{Pratt2009, Eckmiller2011, Maughan2012, Lovisari2015, Kettula2015, Zou2016, Giles2016, Bahar2022}.

In this work, we fit the $L_{\rm X}-T$ relation for the three samples namely, ``isolated'', ``surroundings'', and ``lightcone'', using the same statistical framework. By doing that we put constraints on the underlying $L_{\rm X}-T$ relations of the Magneticum simulations for these three samples by fully simulating the observation and fitting pipeline as if this is an observational scaling relation calibrations study. We compare the results for different samples with each other in order to quantify the impact of the surrounding and in-projection structure on the $L_{\rm X}-T$ relation. Moreover, we compare our best-fit relations with the previously reported $L_{\rm X}-T$ scaling relation results to quantify where $L_{\rm X}-T$ relation of \textit{eROSITA}-like observables of Magneticum clusters lie compared to other observational and simulation findings.

We followed a Bayesian approach in order to fit the relations for the three samples. In the fitting of the three samples, we used the same statistical framework where we fully take into account the selection effects and the mass function. The Bayesian framework we employed here is a modified version of the one used in \citet{Bahar2022}.

The cluster sample used in this work is selected by applying a mass cut of $M_{500c} > 10^{14} M_\odot/h$ (see Sect.~\ref{sec:methods_sample}). This is different from the usual X-ray-selected samples that are most commonly used for calibrating the $L_{\rm X}-T$ relation such as the eFEDS sample \citet{Liu2022a,Bahar2022} or the XXL sample \citet{Pacaud2016,Giles2016}. In our framework, the effect of sample selection is taken into account by jointly modeling the $L_{\rm X}-T$ and $T-M_{500c}$ relations, with priors on the $T-M_{500c}$ relation, and marginalizing over the selection observable $M_{500c}$.

The statistical description of this framework is as follows. The joint probability function as a function of the observed ($\hat{L}_{\rm X}$, $\hat{T}$) and true ($L_{\rm X}$, $T$, $M_{500c}$) observables is given by

\begin{equation}
\begin{split}
P(\hat{L}_{\rm X},\hat{T}, L_{\rm X}, T, M_{500c},I| \theta,z) = \ &P(I|M_{500c}, z)P(\hat{L}_{\rm X},\hat{T} | L_{\rm X}, T) \\ & \times P (L_{\rm X}, T|M_{500c}, \theta, z) P(M_{500c} | z),
\end{split}
\label{eqn:joint_pdf}
\end{equation}
where $P(I|M_{500c}, z)$ is the selection function which is defined as the probability of the cluster with a given $M_{500c}$ and $z$ being included ($I$) in the cluster sample, $P(\hat{L}_{\rm X},\hat{T} | L_{\rm X}, T)$ is the probability distribution of the measurement uncertainties of the $L$ and $T$ observables including the covariance between them, $P (L_{\rm X}, T|M_{500c}, \theta, z)$ is the modelled scaling relation between $L_{\rm X}-T$ and $T-M_{500c}$ with free parameters $\theta$, and $P(M_{500c} | z)$ is the mass function.

We modelled the scaling relation term, $P (L_{\rm X}, T|M_{500c}, \theta, z)$, as a bivariate normal distribution in the logarithmic $L_{\rm X}-T-M$ space as 
\begin{equation} \label{eqn:scaling_term_general_form}
P (\log(L_{\rm X}), \log(T)|\log(M_{500c}), \theta, z) = \mathcal{N} (\boldsymbol{\mu},\boldsymbol{\Sigma})
,\end{equation}
where the mean is


\begin{equation}
\label{eqn:mu}
\boldsymbol{\mu} =  \left[ \begin{array}{c}
\log(L_{\rm X}) = \log \left ( A_{\rm lt}~L_{\rm X, piv} \left ( \dfrac{T}{T_{\rm piv}} \right )^{B_{\rm lt}} \left ( \dfrac{E(z)}{E(z_{\rm piv})} \right )^{C_{\rm lt}}  \right ) \\ [3ex]
\log(T) = \log \left ( A_{\rm tm}~T_{\rm piv} \left ( \dfrac{M_{500c}}{M_{500c, \rm  piv}} \right )^{B_{\rm tm}} \left ( \dfrac{E(z)}{E(z_{\rm piv})} \right )^{C_{\rm tm}}  \right ) 
\end{array} \right]
,\end{equation}
and the covariance matrix is
\begin{equation}
\label{eqn:Sigma}
\boldsymbol{\Sigma} =  \left[ \begin{array}{cc}
\sigma_{L_{\rm X}|T}^2 & \rho~\sigma_{L_{\rm X}|T}~\sigma_{T|M_{\rm 500c}} \\ [2ex]
\rho~\sigma_{L_{\rm X}|T}~\sigma_{T|M_{\rm 500c}} & \sigma_{T|M_{\rm 500c}}^2 
\end{array} \right]
.\end{equation}
$\theta$ in Eqn.~\ref{eqn:scaling_term_general_form} includes all the 9 free parameters in Eqn.~\ref{eqn:mu} and Eqn.~\ref{eqn:Sigma} which are, $A_{\rm lt}$, $B_{\rm lt}$, $C_{\rm lt}$, $A_{\rm tm}$, $B_{\rm tm}$, $C_{\rm tm}$, $\sigma_{L_{\rm X}|T}$, $\rho$ and $\sigma_{T|M_{\rm 500c}}$. $L_{\rm X, piv}$, $T_{\rm piv}$, $M_{500c, {\rm piv}}$ and $z_{\rm piv}$ parameters in Eqn.~\ref{eqn:mu} are the pivot values of the corresponding observables. For  $L_{\rm X, piv}$ and $T_{\rm piv}$, we used the median of the measured values of the sample as the pivot value. For $M_{500c, {\rm piv}}$ and $z_{\rm piv}$ we used the median true value that we obtained from the simulation. The summary of the pivot values used in this work is provided in Table~\ref{tab:pivots}.

With our mass selected sample, selection function term, $P(I|M_{500c}, z)$, in Eqn.~\ref{eqn:joint_pdf} simply becomes a unit step function which can be formulated as

\begin{equation}
\label{eqn:selection_function}
P(I|M_{500c}, z) = H (M_{500c}) = 
    \begin{cases}
        0 & M_{500c}\leq 10^{14} M_\odot/h \\
        1 & M_{500c} > 10^{14} M_\odot/h \\

    \end{cases}
.\end{equation}

Lastly, for the mass distribution term, $P(M_{500c} | z)$, we used the \citet{Tinker08} mass function. After modeling all the terms in the joint probability density function, we marginalize over the Eqn.~\ref{eqn:joint_pdf} nuisance variables ($L_{\rm X}$, $T$, $M_{500c}$) in order to get the likelihood of the measured observables ($\hat{L}_{\rm X}$, $\hat{T}$). This gives us an initial likelihood for a single cluster of the form

\begin{equation}
\begin{split}
P(\hat{L}_{\rm X},\hat{T},I| \theta,z) = & \int \int \int_{L_{\rm X}, T, M_{500c}} P(I|M_{500c}, z) P(\hat{L}_{\rm X},\hat{T} | L_{\rm X}, T) \\ & \times P (L_{\rm X}, T|M_{500c}, \theta, z) P(M_{500c} | z) dL_{\rm X} dT dM_{500c}
\end{split}
\label{eqn:likelihood_initial_form}
\end{equation}

After having the initial form of the likelihood we use the Bayes theorem to get the conditional likelihood of having $\hat{L}_{{\rm X},i}$ and $\hat{T}_i$ measurements given that the cluster is detected, it is at a redshift of $z_i$ and the trial scaling relations parameters are $\theta$. This gives us the final likelihood for a single cluster of the form

\begin{equation}
\begin{split}
\mathcal{L}(\hat{L}_{{\rm X},i},\hat{T}_i|I, \theta,z_i) & =  \dfrac{P(\hat{L}_{{\rm X},i},\hat{T}_i,I| \theta,z_i) }{\int \int P(\hat{L}_{{\rm X},i},\hat{T}_i,I| \theta,z_i) d\hat{L}_{{\rm X},i}d\hat{T}_i}
\\ & =  \dfrac{P(\hat{L}_{{\rm X},i},\hat{T}_i,I| \theta,z_i) }{P(I|z_i)}
\end{split}
.\label{eqn:likelihood_final_form}
\end{equation}

Lastly, we multiply the final form of the likelihood for each cluster to get the overall likelihood of the sample. This gives us a likelihood of the form

\begin{equation}\label{eqn:likelihood_final_form_sample}
\mathcal{L}(\hat{L}_{\rm X, all},\hat{T}_{\rm all}|I, \theta,z) = \prod_{i}^{\hat{N}_{\rm det}} \mathcal{L}(\hat{L}_{{\rm X},i},\hat{T}_i|I, \theta,z_i)
,\end{equation}
where $\hat{L}_{\rm X, all}$ and $\hat{T}_{\rm all}$ are the measured values of the $L_{\rm X}$ and $T$ observables for all clusters and $\hat{N}_{\rm det}$ is the number of clusters in our sample. 

We note that the denominator in Eqn.~\ref{eqn:likelihood_final_form} does not depend on the model parameters ($\theta$) and therefore is a constant in our bayesian framework. For this reason, one does not need to calculate it over and over again for each likelihood iteration.

\begin{table}
\centering
\caption{List of priors used for fitting the $L_{\rm X}-T$ relations}
\label{tab:priors}
\begin{tabular}{cl}
\hline
\hline
Free parameter   &  Prior  \\
\hline

$A_{\rm lt}$ & $\mathcal{U}(-4,4)$\\
$B_{\rm lt}$ & $\mathcal{U}(-10,10)$ \\
$C_{\rm lt}$ & $\mathcal{U}(-10,10)$\\
$A_{\rm tm}$ & $\mathcal{N}(1.45, 0.14)$\\
$B_{\rm tm}$ & $\mathcal{N}(0.65, 0.11)$\\
$C_{\rm tm}$ & $\mathcal{N}(0.66, 1)$\\
$\sigma_{L_{\rm X}|T}$ & $\mathcal{U}(10^{-4},10)$ \\
$\rho$ & $\mathcal{U}(-10^{-3},10^{-3})$ \\
$\sigma_{T|M_{\rm 500c}}$  & $\mathcal{N}(0.2, 0.023)$\\
\hline
\end{tabular}
\end{table}



\begin{table}
\centering
\caption{Median values of observables measured for the three samples.}
\label{tab:pivots}
\begin{tabular}{cl}
\hline
\hline
Parameters   &  Median/Pivots  \\
\hline
$L_{X, {\rm lightcone}}$ & $ 4 \times 10^{43}~{\rm erg}~{\rm s}^{-1}$ \\
$L_{X, {\rm slice}}$ & $ 3.95 \times 10^{43}~{\rm erg}~{\rm s}^{-1}$ \\
$L_{X, {\rm isolated}}$ & $3.99 \times 10^{43}~{\rm erg}~{\rm s}^{-1}$ \\
$T_{\rm lightcone}$ &  2.26 keV\\ 
$T_{\rm slice}$ &  2.29 keV\\ 
$T_{\rm isolated}$ &  2.29 keV\\ 

$M_{\rm 500c}$ & $3 \times 10^{14}~$M$_{\odot}$\\
$z$ & 0.15 \\
\hline
\end{tabular}
\tablefoot{The values listed here are used as pivot values of observables in Eq.~\ref{eqn:mu}}
\end{table}

We fit all of the three $L_{\rm X}-T$ relations, ``isolated'', ``surroundings'', and ``lightcone'', one for each sample, using this likelihood. We sampled the likelihood using the MCMC sampler package {\tt emcee} \citep{emcee} where we used flat priors for the $A_{\rm lt}$, $B_{\rm lt}$, $C_{\rm lt}$, $\sigma_{L_{\rm X}|T}$ parameters and Gaussian priors for the $A_{\rm tm}$, $B_{\rm tm}$, $C_{\rm tm}$, in the shape of the posterior distributions obtained in \citet{Chiu2022} for the $T-M_{500c}$ relation and a tight Gaussian prior around 0.2 for $\sigma_{T|M_{\rm 500c}}$ that is the intrinsic scatter value of the simulated clusters in our sample. For the priors of the $T-M_{500c}$ relation, we used observationally calibrated \citet{Chiu2022} results rather than the intrinsic $T-M_{500c}$ relation of the clusters in Magneticum simulation. By doing that, we aimed to emulate the real-life scenario the best we can, where the universe is observed with eROSITA without having access to the intrinsic $T-M_{500c}$ relation from the simulation. The list of priors for each free parameter is provided in Table~\ref{tab:priors}.

\subsubsection{L-T Relation: Results and Comparison with Observations}

\begin{figure*}
\centering
\includegraphics[width=0.95\textwidth]{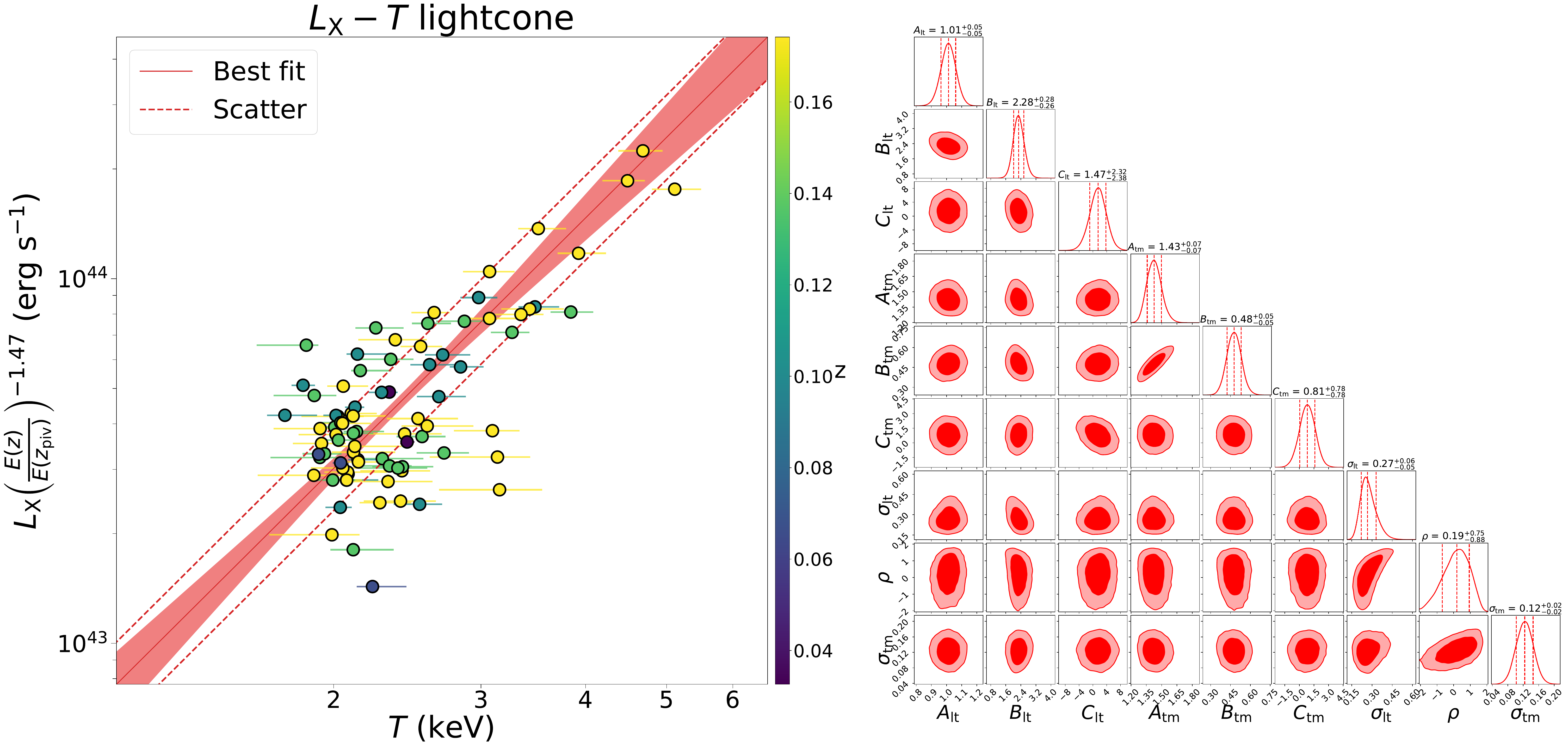}
\caption{{\it Left:} Best fitting scaling relation model to the soft band ($0.5-2.0$ keV) X-ray luminosity ($L_{\rm X}$), temperature ($T$) and redshift ($z$) measurements of the lightcone sample. The solid red line represents the best-fit line, the light-red shaded area represents $\pm 1\sigma$ uncertainty of the mean of the relation (see the first row of $\mu$ in Eq.~\ref{eqn:mu}), and the dashed red line represents the intrinsic scatter of the relation ($\sigma_{L|T}$) around the mean. {\it Right:} Marginal and joint posterior distributions of the jointly modeled $L_{\rm X}-T$ and $T-M_{\rm 500c}$ relations obtained from the second half of the MCMC chains. Red dashed vertical lines indicate the 32nd, 50th, and 68th percentiles and contours indicate 68\% and 95\% credibility regions.
\label{fig:figLT}}
\end{figure*}

As a result of our Bayesian fitting procedure, for each sample, we obtained posterior distributions for the 9 free parameters. Hereby we present best-fit scaling relation parameters of Magneticum clusters measured through an eROSITA-like pipeline in Table~\ref{tab:scalingresults}. We do not observe large variations between the results of different samples. This is expected since the measurement differences between the samples are not very large compared to the error bars. This results in the measurement differences between different samples having a mild effect on the final best-fit values that is taken into account as intrinsic scatter of the relation. Lightcone sample includes the full X-ray emission in projection therefore the observables measured for the lightcone sample are the ones that are the closest to the actual eROSITA measurements (see Sect.~\ref{sec:methods_phox}). Accordingly, for comparison with the literature, the best-fit results of the lightcone sample should be used. The best fitting scaling relation model to the $L_{\rm X}$ and $T$ measurements of the lightcone sample and the posterior distribution of the parameters can be found in Fig.~\ref{fig:figLT}. Self-similar prediction for the $L_{\rm X}-T$ relation is $L_{\rm X} \propto T^{3/2} E(z)$ and our best fitting slope, $B_{\rm lt} = 2.28^{+0.28}_{-0.26}$, is in $3\sigma$ tension with the self-similar prediction. Since advanced X-ray instruments enabled measuring luminosity and temperature of clusters large enough to form statistical samples, a similar tension has been reported in many independent studies \citep[e.g.,][]{Pratt2009, Eckmiller2011, Maughan2012, Hilton2012, Lovisari2015, Zou2016, Giles2016, Bahar2022}. Tension with the self-similar model is expected to emerge if one or more assumptions of the \citet{Kaiser1986} model are violated. The usual suspect for this violation is the self-similar model not including non-gravitational feedback mechanisms such as AGN feedback. Both $L_{\rm X}$ and $T$ are vulnerable to such baryonic processes therefore the change of slope of the $L_{\rm X}-T$ relation compared to the self-similar prediction is governed by the complex relationship between the non-gravitational mechanisms and their effects on these observables.

Furthermore, the best-fit value of the slope we found in this work ($B_{\rm lt} = 2.28^{+0.28}_{-0.26}$) is broadly consistent but slightly shallower than the most recent studies in the literature where the selection effects are taken into account in a sophisticated manner \citep{Lovisari2015, Giles2016, Bahar2022}. Our results for the slope lies within $2\sigma$ statistical uncertainty with the results presented in \citet{Bahar2022} ($B_{\rm lt} = 2.89^{+0.14}_{-0.13}$) and within $1.3\sigma$ statistical uncertainty with the results presented in \citet{Giles2016} ($B_{\rm lt} = 2.63 \pm 0.15$) and \citet{Lovisari2015} ($B_{\rm lt} = 2.67 \pm 0.11$). We argue that the origin of the slope mismatch may be due to two reasons. The first possible cause is investigating $L_{\rm X}-T$ relation using samples  living in different mass parameter spaces may be leading to slightly different results. The cluster sample used in this work is obtained by applying a mass cut of $>10^{14} M_{\odot}$ which results in the sample being made up of mostly $M_{500c} \sim 10^{14} M_{\odot}$ clusters because of the steep mass function. However, for example, the sample used in \citet{Bahar2022} contains a significant amount of galaxy groups that cover $<10^{14} M_{\odot}$ parameter space. In fact, \citet{Lovisari2015} found galaxy groups having a steeper $L_{\rm X}-T$ relation when they compared their results obtained for their galaxy groups sample ($B_{\rm lt} = 2.86 \pm 0.29$) and their high mass sample ($B_{\rm lt} = 2.55 \pm 0.27$). The second possible reason for the slight mismatch is the implementation of non-gravitational processes in simulations being challenging that alter these relations. Recently there has been a significant improvement in implementing non-gravitational feedback mechanisms in simulations however it is an open question whether the modeling is accurate enough to study the relation, $L_{\rm X}-T$, that is arguably affected the most by these mechanisms.

Besides having broadly consistent findings with the recently reported results that fully take into account the selection effects, our best-fitting slope is also consistent with other results in the literature. The slope reported in \citet{Kettula2015} ($B_{\rm lt} = 2.52 \pm 0.10$) and \citet{Eckmiller2011} ($B_{\rm lt} = 2.52 \pm 0.17$) are also slightly steeper but in very good agreement with our results. \citet{Pratt2009} reported a slope of $B_{\rm lt} = 2.24 \pm 0.25$ which is very close to our results whereas the error bar of their measurement is relatively large. \citet{Biffi2013} found a slope of $B_{\rm lt} = 1.97 \pm 0.23$ for a smaller set of clusters from a lower-resolution version of the Magneticum simulation used in this work. \citet{Biffi2014} studied the same relation using Marenostrum MUltidark SImulations of galaxy Clusters (MUSIC) data set and found a slope of $B_{\rm lt} = 2.24 \pm 0.25$ when they used BCES bisector (Y, X) method that is also in good agreement with our findings. Our cluster sample covers a redshift range of 0.03-0.17 which is relatively small compared to the redshift span of other samples used in observational studies. This results in our best-fit redshift evolution parameter being unconstrained $C_{\rm lt} = 1.47^{+2.32}_{-2.38} $. We note that even if the redshift evolution parameter cannot be constrained, it is better to have it as a free parameter in order to have the most realistic statistical uncertainties possible on other parameters. For intrinsic scatter of the $L_{\rm X}-T$ relation, we found a best-fit value of $\sigma_{L_{\rm X}|T} = 0.27^{+0.06}_{-0.05}$. Our finding is considerably smaller than the previously reported results by \citet{Bahar2022} ($\sigma_{L_{\rm X}|T} = 0.78^{+0.08}_{-0.07}$) and \citet{Pratt2009} ($\sigma_{L_{\rm X}|T} = 0.76 \pm 0.14$) whereas still smaller but $~2.2 \sigma$ away from the scatter reported in \citet{Giles2016} ($\sigma_{L_{\rm X}|T} = 0.47 \pm 0.07$). Finding a smaller intrinsic scatter could be due to insufficient modeling at various steps in both observational measurements and simulations. On the observation side, any observational or physical fluctuation that is not modeled other than the physical intrinsic scatter of clusters will add to the intrinsic scatter, and on the simulation side, any observational or physical fluctuation that is missing or under-estimated in the photon simulations will result in having low scatter. Linking the scatter in simulations and observations exceeds the scope of this work therefore we leave the investigation to future work.

We note that there are no mass measurements included in our Bayesian fitting framework. The $T-M_{\rm 500c}$ relation and the mass integral are included only to robustly model the mass-dependent cluster selection. Therefore constraining the $T-M_{\rm 500c}$ relation is not among the primary goals of this work. The sampling distribution of the $A_{\rm tm}$, $B_{\rm tm}$, $C_{\rm tm}$ parameters are mostly driven by their priors. The scattered $T-M_{\rm 500c}$ relation modeled in this work only has an impact on the modeled $L_{\rm X}-T$ distribution through the lowest $\hat{L}_{\rm X}$, $\hat{T}$ measurements where the mass-dependent selection has the most prominent effect on the likelihood along the $L_{\rm X}-T$ plane. As a result, not surprisingly the best-fit values of the $A_{\rm tm}$, $B_{\rm tm}$, $C_{\rm tm}$ parameters for all samples are within the $2\sigma$ confidence region of the prior distribution. Lastly, we rerun our fitting pipeline with uniform priors on the scaling relation parameters of the $T-M_{\rm 500c}$ relation in order to investigate the impact of the priors on the best-fit parameters of the $L_{\rm X}-T$ relation. We find the results for the $L_{\rm X}-T$ relations are all well within $1\sigma$ statistical uncertainty of the results with Gaussian priors, whereas as one would expect, the chains for the $T-M_{\rm 500c}$ relation parameters wander around in unrealistic parameter space due to the lack of $T$ and $M_{\rm 500c}$ measurements.

\begin{table*}
\centering
\caption{Best-fit parameters of the $L_{\rm X}-T$ and $T-M_{\rm 500c}$ relations for different samples}
\renewcommand{\arraystretch}{1.4}
\begin{tabular}{ l c c c c c c c c c c }
\hline
\hline


Sample & $A_{\rm lt}$ & $B_{\rm lt}$ & $C_{\rm lt}$  & $A_{\rm tm}$ & $B_{\rm tm}$ & $C_{\rm tm}$ & $\sigma_{L_{\rm X}|T}$  & $\rho$ & $\sigma_{T|M_{\rm 500c}}$ \\
\hline


Lightcone & $1.01^{+0.05}_{-0.05}$ & $2.28^{+0.28}_{-0.26}$ & $1.47^{+2.32}_{-2.38}$ & $1.43^{+0.07}_{-0.07}$ & $0.48^{+0.05}_{-0.05}$ & $0.81^{+0.78}_{-0.78}$ & $0.27^{+0.06}_{-0.05}$ & $0.19^{+0.75}_{-0.88}$ & $0.12^{+0.02}_{-0.02}$ \\
Surroundings & $1.03^{+0.05}_{-0.05}$ & $2.26^{+0.29}_{-0.26}$ & $1.18^{+2.34}_{-2.54}$ & $1.41^{+0.07}_{-0.06}$ & $0.48^{+0.05}_{-0.05}$ & $0.87^{+0.78}_{-0.79}$ & $0.28^{+0.06}_{-0.05}$ & $0.24^{+0.71}_{-0.84}$ & $0.13^{+0.02}_{-0.02}$ \\ 
Isolated & $1.02^{+0.05}_{-0.05}$ & $2.29^{+0.29}_{-0.27}$ & $1.40^{+2.46}_{-2.50}$ & $1.42^{+0.07}_{-0.06}$ & $0.48^{+0.06}_{-0.05}$ & $0.78^{+0.80}_{-0.79}$ & $0.27^{+0.06}_{-0.05}$ & $0.20^{+0.72}_{-0.88}$ & $0.13^{+0.02}_{-0.02}$ \\  

\hline

\end{tabular}
\label{tab:scalingresults}
\tablefoot{Fitted relations are of the forms $L_{\rm X} = A_{\rm lt}~L_{\rm X, piv} \left ( \dfrac{T}{T_{\rm piv}} \right )^{B_{\rm lt}} \left ( \dfrac{E(z)}{E(z_{\rm piv})} \right )^{C_{\rm lt}}  $ and $T = A_{\rm tm}~T_{\rm piv} \left ( \dfrac{M_{500c}}{M_{500c, \rm  piv}} \right )^{B_{\rm tm}} \left ( \dfrac{E(z)}{E(z_{\rm piv})} \right )^{C_{\rm tm}}$ with log-normal intrinsic scatters $\sigma_{L_{\rm X}|T}$ and $\sigma_{T|M_{\rm 500c}}$ (in natural log). Pivot values used in these relations are provided in Table~\ref{tab:pivots}. These relations are connected to each other via the common observable $T$ and a cross-correlation parameter $\rho$. Detailed description of the joint modeling and fitting of these relations as a multivariate normal distribution in log-log-log $L_{\rm X}-T-M$ space can be found in Sect.~\ref{sec:lt_relation}.
Errors provided here are obtained from the second half of the MCMC chains and represent 1$\sigma$ statistical uncertainties.}
\end{table*}

\section{Conclusions}\label{sec:conc}

In this work, we have carried out an analysis of mock \textit{eROSITA} observations of 84 clusters
from the Magneticum ``Box2\_hr'' cosmological simulation, which were processed through the SIXTE simulator. Our conclusions are as follows:

\begin{itemize}
       
\item We first produced simple simulations of thermal spectra with lognormal temperature distributions convolved with the \textit{eROSITA} responses, over a range of central temperatures $\ln{kT_0}$ and spreads $\ln{\sigma}$. We found that for values of $\ln{\sigma} \leq 0.2$, the temperature obtained from single-temperature fits is always $\lesssim 5$\% from the central temperature, but for larger values of $\ln{\sigma}$ the fitted temperature more significantly underestimates the central temperature by $\sim$20-30\% at $kT \sim 1-2$~keV (depending on the redshift and metallicity) and up to $\sim$50\% at larger temperatures ($kT \gtrsim 10$~keV) and much lower temperatures ($kT \lesssim 0.6$~keV). However, these extreme temperatures will not be the focus of cluster studies with \textit{eROSITA}. 

\item We derived a ``spectroscopic-like'' temperature for the clusters in our sample along the lines of \citet{Mazzotta2004}, and determined that a weighting function of $w = n_en_pT^{-\alpha}$ with $\alpha = 0.76 \substack{+0.05 \\ -0.04}$ (assuming a cylindrical region of $r_{500c}$ for computing the weights of the gas particles in the clusters) is the best-fit to our sample, which is consistent with the value from M04. If we compute the weights using all of the gas particles within the spherical region $r_{500c}$, the best-fit $\alpha = 0.85 \substack{+0.05 \\ -0.03}$. The 1-$\sigma$ accuracy of this temperature compared with the fitted temperature is $\sim$8\%, with some differences as much as $\sim$20\%, which is not nearly as accurate as the $T_{\rm sl}$ derived for clusters with $kT \gtrsim 2-3$~keV and \textit{Chandra} and \textit{XMM-Newton} observations by M04. Investigating a way to more accurately predict of single-temperature fits to \textit{eROSITA} spectra from simulations is left for future work. We also compared the fractional difference of the fitted temperature to the log-normal central temperature from the SPH particles to the same quantity from the toy models, and find general agreement. 

\item We also compare the luminosities computed directly from the simulation gas particles to the luminosities estimated from single-temperature fits, both within an overdensity radius of $r_{500c}$. If spheres of $r_{500c}$ are used, the fitted luminosity overestimates the actual luminosity by $\sim$6\% on average, since the former uses emission projected within a cylindrical region along the line of sight within the same radius. If we instead compare to the simulation-derived luminosity within the same cylindrical region, the agreement is improved, though there is still a scatter of $\sim$5\% between the simulation and fitted estimates. This scatter originates from fitting single-temperature and metallicity models to spectra which include emission from gas at various temperatures and metallicities.

\item We compared temperatures and luminosities from three different samples for the 84 clusters, where other structures in projection were progressively added, first near to each cluster at roughly the same redshift, and finally across a lightcone of emission over a range of redshift. We find that the differences in temperature and luminosity between these samples are all very small, with mean differences on the order of $\sim$1-2\% and 1-$\sigma$ scatter of $\sim$2-3\%. The most extreme examples of differences in luminosity and temperature arise from obvious projections of structures external to the main cluster under consideration that may be easily accounted for in analysis.

\item We fitted an $L_{\rm X}-T$ relation to the eROSITA-like measurements for the three different samples following a Bayesian approach by jointly modeling $L_{\rm X}-T$ and $T-M_{\rm 500c}$ relations in order to take into account the selection effects and the mass function. We constrained the $L_{\rm X}-T$ relation through mock observed $\hat{L}_{\rm X}$ and $\hat{T}$ measurements and $T-M_{\rm 500c}$ scaling relation parameters are left free with priors taken from the literature in order to robustly model the selection function. Parallel to the similarities in $L_{\rm X}$ and $T$ measurements, we find the best-fit parameters of the $L_{\rm X}-T$ relation of different samples being practically the same within the error bars. Furthermore, we compared our $L_{\rm X}-T$ scaling relation results with the literature for the lightcone sample which is closest to the eROSITA observations. We found our slope ($B_{\rm lt} = 2.28^{+0.28}_{-0.26}$) being broadly consistent but slightly shallower than the recently reported results that fully account for the selection effects. Given the limited redshift span of our cluster sample, our fitting machinery was unable to constrain the redshift evolution ($C_{\rm lt} = 1.47^{+2.32}_{-2.38}$), however its contribution to the uncertainties of other measurements is included. Compared to the literature, we found a smaller intrinsic scatter ($\sigma_{L_{\rm X}|T} = 0.27^{+0.06}_{-0.05}$) which we argue may indicate insufficient modeling of observational and/or physical variations in observational studies and/or in simulations.

\end{itemize}

Overall, the bias in temperature and luminosity of clusters induced by projection effects from structures outside the system in question is very small for almost all of the clusters in our sample. This bias is smaller than the expected statistical errors from the \textit{eROSITA} observations, systematic differences due to fitting single-temperature models to multi-temperature gas, and the bias induced by using a projected luminosity to estimate one computed within a sphere of the same radius. This indicates that consideration of projection effects from external structures should not be a large concern for studies using observed properties of clusters as mass proxies for constraining cosmological parameters, and the focus should be on differences arising from multiphase gas and geometrical considerations. 

These conclusions necessarily come with some caveats. Our analysis should be extended to larger sample sizes of clusters, corresponding to lightcones with wider angular sizes. With larger sample sizes, chance alignments between clusters along the line of sight will inevitably increase. When studying larger samples, it would also be instructive to examine the effect of varying cosmological parameters on projection effects, especially those parameters which would increase the number of chance alignments between clusters (though for the range of cosmological parameters currently permitted by observations these effects are likely to be small). The most significant projection effects would occur in systems for which our line of sight is aligned by chance with a cosmic filament stretching Mpc in length at the same location as the target cluster on the sky. Constructing mocks from cosmological simulations where such alignments are purposefully chosen could give a ``worst-case'' estimate of projection effects. Finally, in this paper we have used all of the simulated counts from each cluster in the analysis. Given the variation in core properties in clusters, many analyses of scaling relations work instead with core-excised quantities. It would be interesting to perform this same analysis with core-excised quantities, and also investigate the properties of clusters in merging versus relaxed samples. These considerations we leave for future work. 

\begin{acknowledgements}
The Magneticum Pathfinder simulations have been performed at the Leibniz-Rechenzentrum with CPU time
assigned to the projects pr86re and pr83li. JAZ thanks Alexey Vikhlinin and Urmila Chadayamurri for
useful discussions. JAZ acknowledges support from the Chandra X-ray Center, which is operated by the
Smithsonian Astrophysical Observatory for and on behalf of NASA under contract NAS8-03060. EB
acknowledges financial support from the European Research Council (ERC) Consolidator Grant under the
European Union's Horizon 2020 research and innovation programme (grant agreement CoG DarkQuest No
101002585). VB acknowledges support by the Deutsche Forschungsgemeinschaft (DFG, German Research
Foundation) - 415510302. KD acknowledges support by the COMPLEX project from the European Research Council (ERC) under the European Union’s Horizon 2020 research and innovation program grant agreement ERC-2019-AdG 882679 and by the Excellence Cluster ORIGINS which is funded by the DFG under Germany's Excellence Strategy - EXC-2094 - 390783311. TD and OK acknowledge funding by the Deutsches Zentrum für Luft- und Raumfahrt contract 50 QR 2103. Software packages used in this work include: \texttt{PHOX}\footnote{\url{https://www.usm.uni-muenchen.de/~biffi/phox.html}} \citep{Biffi2012,Biffi2013}; \texttt{SIXTE}\footnote{\url{https://www.sternwarte.uni-erlangen.de/sixte/}} \citep{Dauser2019}; AstroPy\footnote{\url{https://www.astropy.org}} \citep{AstroPy2013}; Matplotlib; NumPy\footnote{\url{https://www.numpy.org}} \citep{Harris2020}; yt\footnote{\url{https://yt-project.org}} \citep{Turk2011}

\end{acknowledgements}
       
%
%

\bibliographystyle{aa}
\bibliography{ms}       

\begin{thebibliography}{68}
\expandafter\ifx\csname natexlab\endcsname\relax\def\natexlab#1{#1}\fi

\bibitem[{{Astropy Collaboration} {et~al.}(2013){Astropy Collaboration},
  {Robitaille}, {Tollerud}, {Greenfield}, {Droettboom}, {Bray}, {Aldcroft},
  {Davis}, {Ginsburg}, {Price-Whelan}, {Kerzendorf}, {Conley}, {Crighton},
  {Barbary}, {Muna}, {Ferguson}, {Grollier}, {Parikh}, {Nair}, {Unther},
  {Deil}, {Woillez}, {Conseil}, {Kramer}, {Turner}, {Singer}, {Fox}, {Weaver},
  {Zabalza}, {Edwards}, {Azalee Bostroem}, {Burke}, {Casey}, {Crawford},
  {Dencheva}, {Ely}, {Jenness}, {Labrie}, {Lim}, {Pierfederici}, {Pontzen},
  {Ptak}, {Refsdal}, {Servillat}, \& {Streicher}}]{AstroPy2013}
{Astropy Collaboration}, {Robitaille}, T.~P., {Tollerud}, E.~J., {et~al.} 2013,
  Astronomy \& Astrophysics, 558, A33

\bibitem[{{Bahar} {et~al.}(2022){Bahar}, {Bulbul}, {Clerc}, {Ghirardini},
  {Liu}, {Nandra}, {Pacaud}, {Chiu}, {Comparat}, {Ider-Chitham}, {Klein},
  {Liu}, {Merloni}, {Migkas}, {Okabe}, {Ramos-Ceja}, {Reiprich}, {Sanders}, \&
  {Schrabback}}]{Bahar2022}
{Bahar}, Y.~E., {Bulbul}, E., {Clerc}, N., {et~al.} 2022, \aap, 661, A7

\bibitem[{{Becker} \& {Kravtsov}(2011)}]{Becker2011}
{Becker}, M.~R. \& {Kravtsov}, A.~V. 2011, \apj, 740, 25

\bibitem[{{Biffi} {et~al.}(2013){Biffi}, {Dolag}, \&
  {B{\"o}hringer}}]{Biffi2013}
{Biffi}, V., {Dolag}, K., \& {B{\"o}hringer}, H. 2013, \mnras, 428, 1395

\bibitem[{{Biffi} {et~al.}(2012){Biffi}, {Dolag}, {B{\"o}hringer}, \&
  {Lemson}}]{Biffi2012}
{Biffi}, V., {Dolag}, K., {B{\"o}hringer}, H., \& {Lemson}, G. 2012, \mnras,
  420, 3545

\bibitem[{{Biffi} {et~al.}(2022){Biffi}, {Dolag}, {Reiprich}, {Veronica},
  {Ramos-Ceja}, {Bulbul}, {Ota}, \& {Ghirardini}}]{Biffi2022}
{Biffi}, V., {Dolag}, K., {Reiprich}, T.~H., {et~al.} 2022, \aap, 661, A17

\bibitem[{{Biffi} {et~al.}(2014){Biffi}, {Sembolini}, {De Petris},
  {Valdarnini}, {Yepes}, \& {Gottl{\"o}ber}}]{Biffi2014}
{Biffi}, V., {Sembolini}, F., {De Petris}, M., {et~al.} 2014, \mnras, 439, 588

\bibitem[{{Brunner} {et~al.}(2022){Brunner}, {Liu}, {Lamer}, {Georgakakis},
  {Merloni}, {Brusa}, {Bulbul}, {Dennerl}, {Friedrich}, {Liu}, {Maitra},
  {Nandra}, {Ramos-Ceja}, {Sanders}, {Stewart}, {Boller}, {Buchner}, {Clerc},
  {Comparat}, {Dwelly}, {Eckert}, {Finoguenov}, {Freyberg}, {Ghirardini},
  {Gueguen}, {Haberl}, {Kreykenbohm}, {Krumpe}, {Osterhage}, {Pacaud},
  {Predehl}, {Reiprich}, {Robrade}, {Salvato}, {Santangelo}, {Schrabback},
  {Schwope}, \& {Wilms}}]{Brunner2022}
{Brunner}, H., {Liu}, T., {Lamer}, G., {et~al.} 2022, \aap, 661, A1

\bibitem[{{Bulbul} {et~al.}(2019){Bulbul}, {Chiu}, {Mohr}, {McDonald},
  {Benson}, {Bautz}, {Bayliss}, {Bleem}, {Brodwin}, {Bocquet}, {Capasso},
  {Dietrich}, {Forman}, {Hlavacek-Larrondo}, {Holzapfel}, {Khullar}, {Klein},
  {Kraft}, {Miller}, {Reichardt}, {Saro}, {Sharon}, {Stalder}, {Schrabback}, \&
  {Stanford}}]{Bulbul2019}
{Bulbul}, E., {Chiu}, I.~N., {Mohr}, J.~J., {et~al.} 2019, \apj, 871, 50

\bibitem[{{Bulbul} {et~al.}(2022){Bulbul}, {Liu}, {Pasini}, {Comparat},
  {Hoang}, {Klein}, {Ghirardini}, {Salvato}, {Merloni}, {Seppi}, {Wolf},
  {Anderson}, {Bahar}, {Brusa}, {Br{\"u}ggen}, {Buchner}, {Dwelly},
  {Ibarra-Medel}, {Ider Chitham}, {Liu}, {Nandra}, {Ramos-Ceja}, {Sanders}, \&
  {Shen}}]{Bulbul2022}
{Bulbul}, E., {Liu}, A., {Pasini}, T., {et~al.} 2022, \aap, 661, A10

\bibitem[{{Bulbul} {et~al.}(2010){Bulbul}, {Hasler}, {Bonamente}, \&
  {Joy}}]{Bulbul2010}
{Bulbul}, G.~E., {Hasler}, N., {Bonamente}, M., \& {Joy}, M. 2010, \apj, 720,
  1038

\bibitem[{{Cash}(1979)}]{Cash1979}
{Cash}, W. 1979, \apj, 228, 939

\bibitem[{{Chiu} {et~al.}(2022){Chiu}, {Ghirardini}, {Liu}, {Grandis},
  {Bulbul}, {Bahar}, {Comparat}, {Bocquet}, {Clerc}, {Klein}, {Liu}, {Li},
  {Miyatake}, {Mohr}, {More}, {Oguri}, {Okabe}, {Pacaud}, {Ramos-Ceja},
  {Reiprich}, {Schrabback}, \& {Umetsu}}]{Chiu2022}
{Chiu}, I.~N., {Ghirardini}, V., {Liu}, A., {et~al.} 2022, \aap, 661, A11

\bibitem[{{Dauser} {et~al.}(2019){Dauser}, {Falkner}, {Lorenz}, {Kirsch},
  {Peille}, {Cucchetti}, {Schmid}, {Brand}, {Oertel}, {Smith}, \&
  {Wilms}}]{Dauser2019}
{Dauser}, T., {Falkner}, S., {Lorenz}, M., {et~al.} 2019, \aap, 630, A66

\bibitem[{{Davis} {et~al.}(1985){Davis}, {Efstathiou}, {Frenk}, \&
  {White}}]{Davis1985}
{Davis}, M., {Efstathiou}, G., {Frenk}, C.~S., \& {White}, S.~D.~M. 1985, \apj,
  292, 371

\bibitem[{{Dolag} {et~al.}(2009){Dolag}, {Borgani}, {Murante}, \&
  {Springel}}]{Dolag2009}
{Dolag}, K., {Borgani}, S., {Murante}, G., \& {Springel}, V. 2009, \mnras, 399,
  497

\bibitem[{{Dolag} {et~al.}(2015){Dolag}, {Gaensler}, {Beck}, \&
  {Beck}}]{2015MNRAS.451.4277D}
{Dolag}, K., {Gaensler}, B.~M., {Beck}, A.~M., \& {Beck}, M.~C. 2015, \mnras,
  451, 4277

\bibitem[{{Dolag} {et~al.}(2006){Dolag}, {Meneghetti}, {Moscardini}, {Rasia},
  \& {Bonaldi}}]{Dolag2006}
{Dolag}, K., {Meneghetti}, M., {Moscardini}, L., {Rasia}, E., \& {Bonaldi}, A.
  2006, \mnras, 370, 656

\bibitem[{{Dolag} {et~al.}(2017){Dolag}, {Mevius}, \& {Remus}}]{Dolag2017}
{Dolag}, K., {Mevius}, E., \& {Remus}, R.-S. 2017, Galaxies, 5, 35

\bibitem[{{Eckmiller} {et~al.}(2011){Eckmiller}, {Hudson}, \&
  {Reiprich}}]{Eckmiller2011}
{Eckmiller}, H.~J., {Hudson}, D.~S., \& {Reiprich}, T.~H. 2011, \aap, 535, A105

\bibitem[{{Ettori} {et~al.}(2019){Ettori}, {Ghirardini}, {Eckert},
  {Pointecouteau}, {Gastaldello}, {Sereno}, {Gaspari}, {Ghizzardi},
  {Roncarelli}, \& {Rossetti}}]{Ettori2019}
{Ettori}, S., {Ghirardini}, V., {Eckert}, D., {et~al.} 2019, \aap, 621, A39

\bibitem[{{Ezer} {et~al.}(2017){Ezer}, {Bulbul}, {Nihal Ercan}, {Smith},
  {Bautz}, {Loewenstein}, {McDonald}, \& {Miller}}]{Ezer2017}
{Ezer}, C., {Bulbul}, E., {Nihal Ercan}, E., {et~al.} 2017, \apj, 836, 110

\bibitem[{{Foreman-Mackey} {et~al.}(2013){Foreman-Mackey}, {Hogg}, {Lang}, \&
  {Goodman}}]{emcee}
{Foreman-Mackey}, D., {Hogg}, D.~W., {Lang}, D., \& {Goodman}, J. 2013, \pasp,
  125, 306

\bibitem[{{Frank} {et~al.}(2013){Frank}, {Peterson}, {Andersson}, {Fabian}, \&
  {Sanders}}]{Frank2013}
{Frank}, K.~A., {Peterson}, J.~R., {Andersson}, K., {Fabian}, A.~C., \&
  {Sanders}, J.~S. 2013, \apj, 764, 46

\bibitem[{{Gianfagna} {et~al.}(2023){Gianfagna}, {Rasia}, {Cui}, {De Petris},
  {Yepes}, {Contreras-Santos}, \& {Knebe}}]{Gianfagna2023}
{Gianfagna}, G., {Rasia}, E., {Cui}, W., {et~al.} 2023, \mnras, 518, 4238

\bibitem[{{Giles} {et~al.}(2016){Giles}, {Maughan}, {Pacaud}, {Lieu}, {Clerc},
  {Pierre}, {Adami}, {Chiappetti}, {D{\'e}mocl{\'e}s}, {Ettori}, {Le
  F{\'e}vre}, {Ponman}, {Sadibekova}, {Smith}, {Willis}, \&
  {Ziparo}}]{Giles2016}
{Giles}, P.~A., {Maughan}, B.~J., {Pacaud}, F., {et~al.} 2016, \aap, 592, A3

\bibitem[{Harris {et~al.}(2020)Harris, Millman, van~der Walt, Gommers,
  Virtanen, Cournapeau, Wieser, Taylor, Berg, Smith, Kern, Picus, Hoyer, van
  Kerkwijk, Brett, Haldane, del R{\'{i}}o, Wiebe, Peterson,
  G{\'{e}}rard-Marchant, Sheppard, Reddy, Weckesser, Abbasi, Gohlke, \&
  Oliphant}]{Harris2020}
Harris, C.~R., Millman, K.~J., van~der Walt, S.~J., {et~al.} 2020, Nature, 585,
  357

\bibitem[{{Hilton} {et~al.}(2012){Hilton}, {Romer}, {Kay}, {Mehrtens},
  {Lloyd-Davies}, {Thomas}, {Short}, {Mayers}, {Rooney}, {Stott}, {Collins},
  {Harrison}, {Hoyle}, {Liddle}, {Mann}, {Miller}, {Sahl{\'e}n}, {Viana},
  {Davidson}, {Hosmer}, {Nichol}, {Sabirli}, {Stanford}, \&
  {West}}]{Hilton2012}
{Hilton}, M., {Romer}, A.~K., {Kay}, S.~T., {et~al.} 2012, \mnras, 424, 2086

\bibitem[{{Hirschmann} {et~al.}(2014){Hirschmann}, {Dolag}, {Saro}, {Bachmann},
  {Borgani}, \& {Burkert}}]{Hirschmann2014}
{Hirschmann}, M., {Dolag}, K., {Saro}, A., {et~al.} 2014, \mnras, 442, 2304

\bibitem[{{Iljenkarevic} {et~al.}(2022){Iljenkarevic}, {Reiprich}, {Pacaud},
  {Veronica}, {Whelan}, {Aschersleben}, {Migkas}, {Bulbul}, {Sanders},
  {Ramos-Ceja}, {Liu}, {Ghirardini}, {Liu}, \& {Boller}}]{Iljenkarevic2022}
{Iljenkarevic}, J., {Reiprich}, T.~H., {Pacaud}, F., {et~al.} 2022, \aap, 661

\bibitem[{{Kaastra}(2017)}]{Kaastra2017}
{Kaastra}, J.~S. 2017, \aap, 605, A51

\bibitem[{{Kaastra} {et~al.}(2004){Kaastra}, {Tamura}, {Peterson}, {Bleeker},
  {Ferrigno}, {Kahn}, {Paerels}, {Piffaretti}, {Branduardi-Raymont}, \&
  {B{\"o}hringer}}]{Kaastra2004}
{Kaastra}, J.~S., {Tamura}, T., {Peterson}, J.~R., {et~al.} 2004, \aap, 413,
  415

\bibitem[{{Kaiser}(1986)}]{Kaiser1986}
{Kaiser}, N. 1986, \mnras, 222, 323

\bibitem[{{Kettula} {et~al.}(2015){Kettula}, {Giodini}, {van Uitert},
  {Hoekstra}, {Finoguenov}, {Lerchster}, {Erben}, {Heymans}, {Hildebrandt},
  {Kitching}, {Mahdavi}, {Mellier}, {Miller}, {Mirkazemi}, {Van Waerbeke},
  {Coupon}, {Egami}, {Fu}, {Hudson}, {Kneib}, {Kuijken}, {McCracken},
  {Pereira}, {Rowe}, {Schrabback}, {Tanaka}, \& {Velander}}]{Kettula2015}
{Kettula}, K., {Giodini}, S., {van Uitert}, E., {et~al.} 2015, \mnras, 451,
  1460

\bibitem[{{Komatsu} {et~al.}(2011){Komatsu}, {Smith}, {Dunkley}, {Bennett},
  {Gold}, {Hinshaw}, {Jarosik}, {Larson}, {Nolta}, {Page}, {Spergel},
  {Halpern}, {Hill}, {Kogut}, {Limon}, {Meyer}, {Odegard}, {Tucker}, {Weiland},
  {Wollack}, \& {Wright}}]{Komatsu2011}
{Komatsu}, E., {Smith}, K.~M., {Dunkley}, J., {et~al.} 2011, \apjs, 192, 18

\bibitem[{{K{\"o}nig} {et~al.}(2022){K{\"o}nig}, {Wilms}, {Arcodia}, {Dauser},
  {Dennerl}, {Doroshenko}, {Haberl}, {H{\"a}mmerich}, {Kirsch}, {Kreykenbohm},
  {Lorenz}, {Malyali}, {Merloni}, {Rau}, {Rauch}, {Sala}, {Schwope},
  {Suleimanov}, {Weber}, \& {Werner}}]{Koenig2022}
{K{\"o}nig}, O., {Wilms}, J., {Arcodia}, R., {et~al.} 2022, \nat, 605, 248

\bibitem[{{Lau} {et~al.}(2011){Lau}, {Nagai}, {Kravtsov}, \&
  {Zentner}}]{Lau2011}
{Lau}, E.~T., {Nagai}, D., {Kravtsov}, A.~V., \& {Zentner}, A.~R. 2011, \apj,
  734, 93

\bibitem[{{Liu} {et~al.}(2022{\natexlab{a}}){Liu}, {Bulbul}, {Ghirardini},
  {Liu}, {Klein}, {Clerc}, {{\"O}zsoy}, {Ramos-Ceja}, {Pacaud}, {Comparat},
  {Okabe}, {Bahar}, {Biffi}, {Brunner}, {Br{\"u}ggen}, {Buchner}, {Ider
  Chitham}, {Chiu}, {Dolag}, {Gatuzz}, {Gonzalez}, {Hoang}, {Lamer}, {Merloni},
  {Nandra}, {Oguri}, {Ota}, {Predehl}, {Reiprich}, {Salvato}, {Schrabback},
  {Sanders}, {Seppi}, \& {Thibaud}}]{Liu2022a}
{Liu}, A., {Bulbul}, E., {Ghirardini}, V., {et~al.} 2022{\natexlab{a}}, \aap,
  661, A2

\bibitem[{{Liu} {et~al.}(2022{\natexlab{b}}){Liu}, {Bulbul}, {Ramos-Ceja},
  {Sanders}, {Ghirardini}, {Bahar}, {Yeung}, {Gatuzz}, {Freyberg}, {Garrel},
  {Zhang}, {Merloni}, \& {Nandra}}]{Liu2022c}
{Liu}, A., {Bulbul}, E., {Ramos-Ceja}, M.~E., {et~al.} 2022{\natexlab{b}},
  arXiv e-prints, arXiv:2210.00633

\bibitem[{{Liu} {et~al.}(2022{\natexlab{c}}){Liu}, {Merloni}, {Comparat},
  {Nandra}, {Sanders}, {Lamer}, {Buchner}, {Dwelly}, {Freyberg}, {Malyali},
  {Georgakakis}, {Salvato}, {Brunner}, {Brusa}, {Klein}, {Ghirardini}, {Clerc},
  {Pacaud}, {Bulbul}, {Liu}, {Schwope}, {Robrade}, {Wilms}, {Dauser},
  {Ramos-Ceja}, {Reiprich}, {Boller}, \& {Wolf}}]{Liu2022b}
{Liu}, T., {Merloni}, A., {Comparat}, J., {et~al.} 2022{\natexlab{c}}, \aap,
  661, A27

\bibitem[{{Lovisari} {et~al.}(2015){Lovisari}, {Reiprich}, \&
  {Schellenberger}}]{Lovisari2015}
{Lovisari}, L., {Reiprich}, T.~H., \& {Schellenberger}, G. 2015, \aap, 573,
  A118

\bibitem[{{Mantz}(2019)}]{Mantz2019}
{Mantz}, A.~B. 2019, \mnras, 485, 4863

\bibitem[{{Mantz} {et~al.}(2018){Mantz}, {Allen}, {Morris}, \& {von der
  Linden}}]{Mantz2018}
{Mantz}, A.~B., {Allen}, S.~W., {Morris}, R.~G., \& {von der Linden}, A. 2018,
  \mnras, 473, 3072

\bibitem[{{Maughan}(2007)}]{Maughan2007}
{Maughan}, B.~J. 2007, \apj, 668, 772

\bibitem[{{Maughan} {et~al.}(2012){Maughan}, {Giles}, {Randall}, {Jones}, \&
  {Forman}}]{Maughan2012}
{Maughan}, B.~J., {Giles}, P.~A., {Randall}, S.~W., {Jones}, C., \& {Forman},
  W.~R. 2012, \mnras, 421, 1583

\bibitem[{{Mazzotta} {et~al.}(2004){Mazzotta}, {Rasia}, {Moscardini}, \&
  {Tormen}}]{Mazzotta2004}
{Mazzotta}, P., {Rasia}, E., {Moscardini}, L., \& {Tormen}, G. 2004, \mnras,
  354, 10

\bibitem[{{McCammon} {et~al.}(2002){McCammon}, {Almy}, {Apodaca}, {Bergmann
  Tiest}, {Cui}, {Deiker}, {Galeazzi}, {Juda}, {Lesser}, {Mihara},
  {Morgenthaler}, {Sanders}, {Zhang}, {Figueroa-Feliciano}, {Kelley},
  {Moseley}, {Mushotzky}, {Porter}, {Stahle}, \& {Szymkowiak}}]{McCammon2002}
{McCammon}, D., {Almy}, R., {Apodaca}, E., {et~al.} 2002, \apj, 576, 188

\bibitem[{{Mernier} \& {Biffi}(2022)}]{Mernier2022}
{Mernier}, F. \& {Biffi}, V. 2022, arXiv e-prints, arXiv:2202.07097

\bibitem[{{Pacaud} {et~al.}(2016){Pacaud}, {Clerc}, {Giles}, {Adami},
  {Sadibekova}, {Pierre}, {Maughan}, {Lieu}, {Le F{\`e}vre}, {Alis}, {Altieri},
  {Ardila}, {Baldry}, {Benoist}, {Birkinshaw}, {Chiappetti},
  {D{\'e}mocl{\`e}s}, {Eckert}, {Evrard}, {Faccioli}, {Gastaldello}, {Guennou},
  {Horellou}, {Iovino}, {Koulouridis}, {Le Brun}, {Lidman}, {Liske},
  {Maurogordato}, {Menanteau}, {Owers}, {Poggianti}, {Pomar{\`e}de}, {Pompei},
  {Ponman}, {Rapetti}, {Reiprich}, {Smith}, {Tuffs}, {Valageas}, {Valtchanov},
  {Willis}, \& {Ziparo}}]{Pacaud2016}
{Pacaud}, F., {Clerc}, N., {Giles}, P.~A., {et~al.} 2016, \aap, 592, A2

\bibitem[{{Peterson} {et~al.}(2003){Peterson}, {Kahn}, {Paerels}, {Kaastra},
  {Tamura}, {Bleeker}, {Ferrigno}, \& {Jernigan}}]{Peterson2003}
{Peterson}, J.~R., {Kahn}, S.~M., {Paerels}, F.~B.~S., {et~al.} 2003, \apj,
  590, 207

\bibitem[{{Pratt} {et~al.}(2019){Pratt}, {Arnaud}, {Biviano}, {Eckert},
  {Ettori}, {Nagai}, {Okabe}, \& {Reiprich}}]{Pratt2019}
{Pratt}, G.~W., {Arnaud}, M., {Biviano}, A., {et~al.} 2019, \ssr, 215, 25

\bibitem[{{Pratt} {et~al.}(2009){Pratt}, {Croston}, {Arnaud}, \&
  {B{\"o}hringer}}]{Pratt2009}
{Pratt}, G.~W., {Croston}, J.~H., {Arnaud}, M., \& {B{\"o}hringer}, H. 2009,
  \aap, 498, 361

\bibitem[{{Predehl} {et~al.}(2021){Predehl}, {Andritschke}, {Arefiev},
  {Babyshkin}, {Batanov}, {Becker}, {B{\"o}hringer}, {Bogomolov}, {Boller},
  {Borm}, {Bornemann}, {Br{\"a}uninger}, {Br{\"u}ggen}, {Brunner}, {Brusa},
  {Bulbul}, {Buntov}, {Burwitz}, {Burkert}, {Clerc}, {Churazov}, {Coutinho},
  {Dauser}, {Dennerl}, {Doroshenko}, {Eder}, {Emberger}, {Eraerds},
  {Finoguenov}, {Freyberg}, {Friedrich}, {Friedrich}, {F{\"u}rmetz},
  {Georgakakis}, {Gilfanov}, {Granato}, {Grossberger}, {Gueguen}, {Gureev},
  {Haberl}, {H{\"a}lker}, {Hartner}, {Hasinger}, {Huber}, {Ji}, {Kienlin},
  {Kink}, {Korotkov}, {Kreykenbohm}, {Lamer}, {Lomakin}, {Lapshov}, {Liu},
  {Maitra}, {Meidinger}, {Menz}, {Merloni}, {Mernik}, {Mican}, {Mohr},
  {M{\"u}ller}, {Nandra}, {Nazarov}, {Pacaud}, {Pavlinsky}, {Perinati},
  {Pfeffermann}, {Pietschner}, {Ramos-Ceja}, {Rau}, {Reiffers}, {Reiprich},
  {Robrade}, {Salvato}, {Sanders}, {Santangelo}, {Sasaki}, {Scheuerle},
  {Schmid}, {Schmitt}, {Schwope}, {Shirshakov}, {Steinmetz}, {Stewart},
  {Str{\"u}der}, {Sunyaev}, {Tenzer}, {Tiedemann}, {Tr{\"u}mper}, {Voron},
  {Weber}, {Wilms}, \& {Yaroshenko}}]{Predehl2021}
{Predehl}, P., {Andritschke}, R., {Arefiev}, V., {et~al.} 2021, \aap, 647, A1

\bibitem[{{Puchwein} {et~al.}(2008){Puchwein}, {Sijacki}, \&
  {Springel}}]{Puchwein2008}
{Puchwein}, E., {Sijacki}, D., \& {Springel}, V. 2008, \apjl, 687, L53

\bibitem[{{Ramos-Ceja} {et~al.}(2022){Ramos-Ceja}, {Oguri}, {Miyazaki},
  {Ghirardini}, {Chiu}, {Okabe}, {Liu}, {Schrabback}, {Akino}, {Bahar},
  {Bulbul}, {Clerc}, {Comparat}, {Grandis}, {Klein}, {Lin}, {Merloni},
  {Mitsuishi}, {Miyatake}, {More}, {Nandra}, {Nishizawa}, {Ota}, {Pacaud},
  {Reiprich}, \& {Sanders}}]{Ramos-Ceja2022}
{Ramos-Ceja}, M.~E., {Oguri}, M., {Miyazaki}, S., {et~al.} 2022, \aap, 661, A14

\bibitem[{{Sanders} {et~al.}(2022){Sanders}, {Biffi}, {Br{\"u}ggen}, {Bulbul},
  {Dennerl}, {Dolag}, {Erben}, {Freyberg}, {Gatuzz}, {Ghirardini}, {Hoang},
  {Klein}, {Liu}, {Merloni}, {Pacaud}, {Ramos-Ceja}, {Reiprich}, \&
  {ZuHone}}]{Sanders2022}
{Sanders}, J.~S., {Biffi}, V., {Br{\"u}ggen}, M., {et~al.} 2022, \aap, 661, A36

\bibitem[{{Scheck} {et~al.}(2022){Scheck}, {Sanders}, {Biffi}, {Dolag},
  {Bulbul}, \& {Liu}}]{Scheck2022}
{Scheck}, D., {Sanders}, J.~S., {Biffi}, V., {et~al.} 2022, arXiv e-prints,
  arXiv:2211.12146

\bibitem[{{Schellenberger} {et~al.}(2015){Schellenberger}, {Reiprich},
  {Lovisari}, {Nevalainen}, \& {David}}]{Schellenberger2015}
{Schellenberger}, G., {Reiprich}, T.~H., {Lovisari}, L., {Nevalainen}, J., \&
  {David}, L. 2015, \aap, 575, A30

\bibitem[{{Smith} {et~al.}(2001){Smith}, {Brickhouse}, {Liedahl}, \&
  {Raymond}}]{Smith2001}
{Smith}, R.~K., {Brickhouse}, N.~S., {Liedahl}, D.~A., \& {Raymond}, J.~C.
  2001, \apjl, 556, L91

\bibitem[{{Springel}(2005)}]{Springel2005}
{Springel}, V. 2005, \mnras, 364, 1105

\bibitem[{{Springel} {et~al.}(2001){Springel}, {White}, {Tormen}, \&
  {Kauffmann}}]{Springel2001}
{Springel}, V., {White}, S. D.~M., {Tormen}, G., \& {Kauffmann}, G. 2001,
  \mnras, 328, 726

\bibitem[{{Sunyaev} {et~al.}(2021){Sunyaev}, {Arefiev}, {Babyshkin},
  {Bogomolov}, {Borisov}, {Buntov}, {Brunner}, {Burenin}, {Churazov},
  {Coutinho}, {Eder}, {Eismont}, {Freyberg}, {Gilfanov}, {Gureyev}, {Hasinger},
  {Khabibullin}, {Kolmykov}, {Komovkin}, {Krivonos}, {Lapshov}, {Levin},
  {Lomakin}, {Lutovinov}, {Medvedev}, {Merloni}, {Mernik}, {Mikhailov},
  {Molodtsov}, {Mzhelsky}, {M{\"u}ller}, {Nandra}, {Nazarov}, {Pavlinsky},
  {Poghodin}, {Predehl}, {Robrade}, {Sazonov}, {Scheuerle}, {Shirshakov},
  {Tkachenko}, \& {Voron}}]{Sunyaev2021}
{Sunyaev}, R., {Arefiev}, V., {Babyshkin}, V., {et~al.} 2021, \aap, 656, A132

\bibitem[{{Tinker} {et~al.}(2008){Tinker}, {Kravtsov}, {Klypin}, {Abazajian},
  {Warren}, {Yepes}, {Gottl{\"o}ber}, \& {Holz}}]{Tinker08}
{Tinker}, J., {Kravtsov}, A.~V., {Klypin}, A., {et~al.} 2008, \apj, 688, 709

\bibitem[{{Turk} {et~al.}(2011){Turk}, {Smith}, {Oishi}, {Skory}, {Skillman},
  {Abel}, \& {Norman}}]{Turk2011}
{Turk}, M.~J., {Smith}, B.~D., {Oishi}, J.~S., {et~al.} 2011, Astrophysical
  Journals, 192, 9

\bibitem[{{Veronica} {et~al.}(2022){Veronica}, {Su}, {Biffi}, {Reiprich},
  {Pacaud}, {Nulsen}, {Kraft}, {Sanders}, {Bogdan}, {Kara}, {Dolag}, {Kerp},
  {Koribalski}, {Erben}, {Bulbul}, {Gatuzz}, {Ghirardini}, {Hopkins}, {Liu},
  {Migkas}, \& {Vernstrom}}]{Veronica2022}
{Veronica}, A., {Su}, Y., {Biffi}, V., {et~al.} 2022, \aap, 661, A46

\bibitem[{{Vikhlinin}(2006)}]{Vikhlinin2006}
{Vikhlinin}, A. 2006, \apj, 640, 710

\bibitem[{{Whelan} {et~al.}(2022){Whelan}, {Veronica}, {Pacaud}, {Reiprich},
  {Bulbul}, {Ramos-Ceja}, {Sanders}, {Aschersleben}, {Iljenkarevic}, {Migkas},
  {Freyberg}, {Dennerl}, {Kara}, {Liu}, {Ghirardini}, \& {Ota}}]{Whelan2022}
{Whelan}, B., {Veronica}, A., {Pacaud}, F., {et~al.} 2022, \aap, 663, A171

\bibitem[{{Zou} {et~al.}(2016){Zou}, {Maughan}, {Giles}, {Vikhlinin}, {Pacaud},
  {Burenin}, \& {Hornstrup}}]{Zou2016}
{Zou}, S., {Maughan}, B.~J., {Giles}, P.~A., {et~al.} 2016, \mnras, 463, 820

\end{thebibliography}

\begin{appendix}

\section{Fitting With and Without Background}\label{sec:bkgnd}

All of the temperatures and luminosities of the clusters reported in Section \ref{sec:results} were
obtained from spectral fits with background components included in the data and model. In this
Section we report the fitted temperatures and luminosities of the ``isolated'' cluster sample
without background and compare to those with background. The fits without background are otherwise
identical to those with background, e.g. the same energy range is used in the fit and the same
source parameters are frozen and thawed.

\begin{figure*}
\centering
\includegraphics[width=0.95\textwidth]{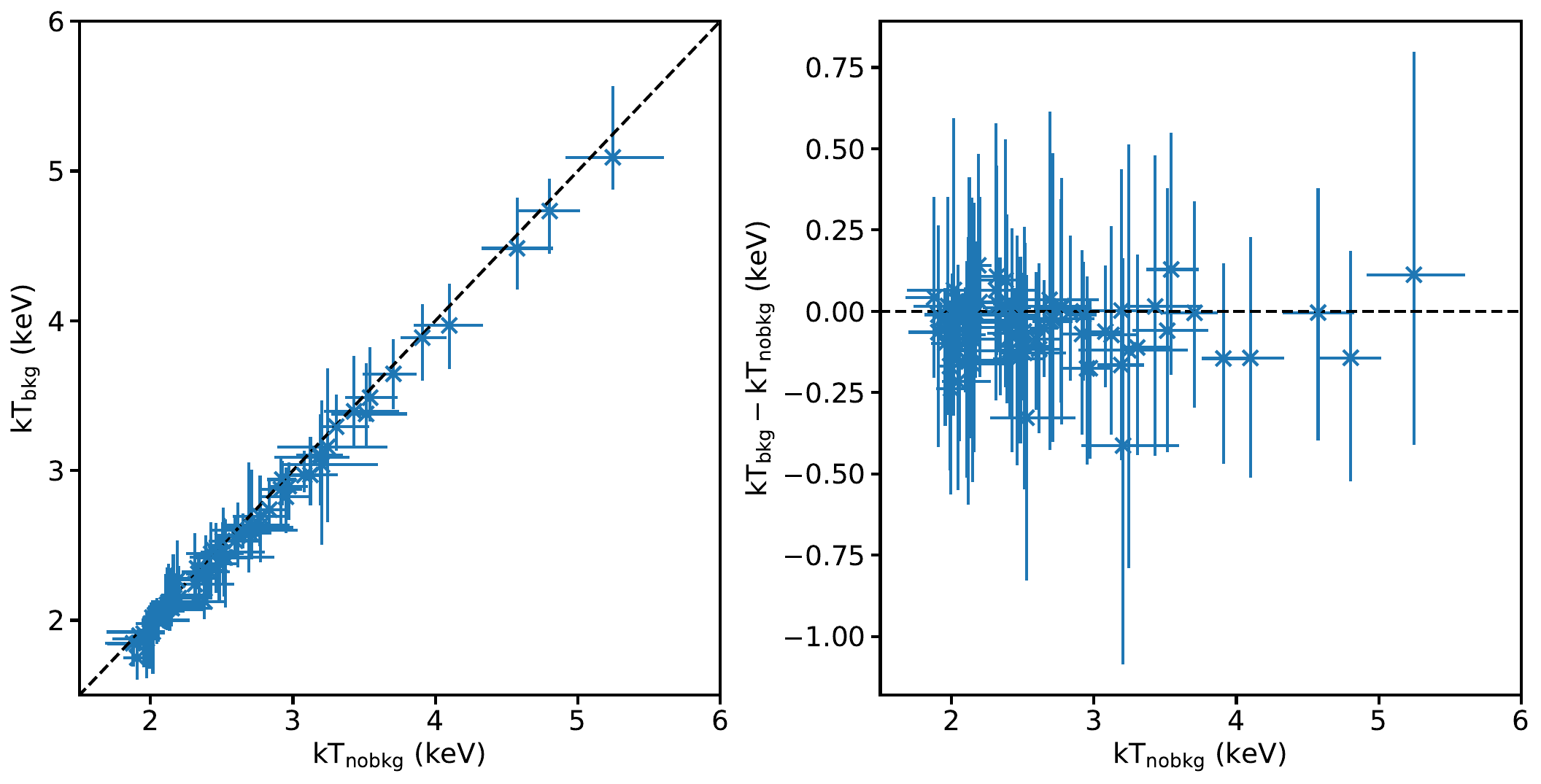}
\caption{Fitted temperatures of the ``isolated'' cluster sample with and without background.
Left panel: the temperature with background vs. the temperature without background (the black dashed
line indicates equality between the two temperatures). Right panel: Difference between the two
temperatures vs. the temperature without background (the black dashed line indicates no difference
between the temperatures).\label{fig:kT_bkg_vs_nobkg}}
\end{figure*}

Figure \ref{fig:kT_bkg_vs_nobkg} shows the fitted temperatures of the ``isolated'' cluster sample
with and without background plotted against each other in the left panel, and with their difference
plotted against the fitted temperature without background in the right panel. The fitted
temperatures with background are typically biased low compared to those without, but the difference
is well within the 1-$\sigma$ errors. Lower-temperature clusters are primarily affected by the astrophysical background and foreground, whereas higher-temperature clusters can be affected also by the non-X-ray background. 

\begin{figure*}
\centering
\includegraphics[width=0.95\textwidth]{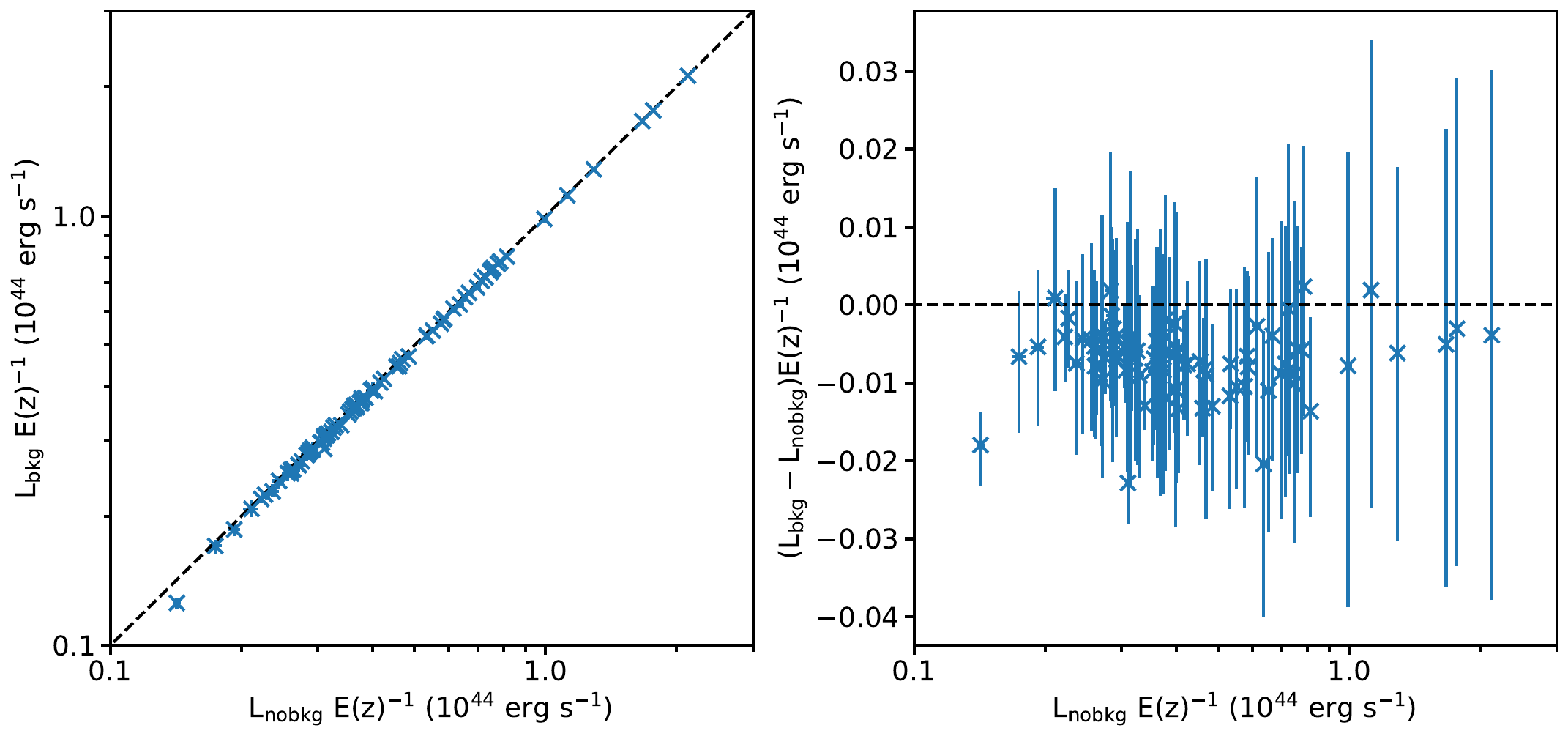}
\caption{Fitted luminosities of the ``isolated'' cluster sample with and without background.
Left panel: the luminosity with background vs. the luminosity without background (the black dashed
line indicates equality between the two luminosities). Right panel: Difference between the two
luminosities vs. the luminosity without background (the black dashed line indicates no difference
between the luminosities).\label{fig:L_bkg_vs_nobkg}}
\end{figure*}

Figure \ref{fig:L_bkg_vs_nobkg} shows the fitted luminosities of the ``isolated'' cluster sample
with and without background plotted against each other in the left panel, and with their difference
plotted against the fitted luminosities without background in the right panel. The fitted
luminosities with background are all biased low compared to those without, but the difference is
very small and almost always within the 1-$\sigma$ error.  

\end{appendix}

\end{document}